\PassOptionsToPackage{tbtags}{amsmath}%This ensures that the equation numbers in displays produced using the ``split'' environment are placed on the final line of the display, as per AIP style (note that for this to work, the ``split'' environment must be use within the ``equation'' environment, not ``eqnarray''.
\documentclass[aip,jcp,reprint,amsmath,amssymb]{revtex4-2}
\usepackage{dcolumn, bm}%the ``indentfirst'' package should not be used for AIP papers, since it incorrectly indents the first line of the abstract
\usepackage{graphicx, tikz, orcidlink}
\usepackage{epstopdf}
\usepackage{ascmac}
\usepackage[normalem]{ulem}%This option ensures that emphasis is indicated by italicization, as per AIP style, rather than underlining
\usepackage{cancel}
\usepackage{here}
\usepackage{multirow}
\usepackage{comment}

\usepackage[]{hyperref}
\usepackage{xcolor}
\definecolor{AIPBlue}{RGB}{61, 180, 229}
\hypersetup{
colorlinks,
linkcolor=AIPBlue,
citecolor=AIPBlue,
urlcolor=AIPBlue
}

\newcounter{aqctr}
\newenvironment{author-query}
{\refstepcounter{aqctr}\par\vspace{\baselineskip}\noindent
\color{red}\textbf{Author Query/Comment AQ \arabic{aqctr}.}}
{\par\vspace{\baselineskip}\normalcolor}

\begin{document}
\title{Classical and quantum thermodynamics described as a system--bath model: The dimensionless minimum work principle}
\date{Last updated: \today}

\author{Shoki Koyanagi \orcidlink{0000-0002-8607-1699}}\email[Authors to whom correspondence should be addressed: ]{koyanagi.syoki.36z@st.kyoto-u.jp and tanimura.yoshitaka.5w@kyoto-u.jp}\affiliation{Department of Chemistry, Graduate School of Science,
Kyoto University, Kyoto 606-8502, Japan}

\author{Yoshitaka Tanimura \orcidlink{0000-0002-7913-054X}}
\email[Authors to whom correspondence should be addressed: ]{koyanagi.syoki.36z@st.kyoto-u.jp and tanimura.yoshitaka.5w@kyoto-u.jp}
\affiliation{Department of Chemistry, Graduate School of Science,
Kyoto University, Kyoto 606-8502, Japan}

\begin{abstract}
We formulate a thermodynamic theory applicable to both classical and quantum systems. These systems are depicted as  thermodynamic system--bath models capable of handling isothermal, isentropic, thermostatic, and entropic processes. Our approach is based on the use of a dimensionless thermodynamic potential expressed as a function of the intensive and extensive thermodynamic variables. 
Using the principles of dimensionless minimum work and dimensionless maximum entropy derived from quasi-static changes of external perturbations and temperature, we obtain the Massieu--Planck potentials as entropic potentials and the Helmholtz--Gibbs potentials as free energy.
These potentials can be interconverted through time-dependent Legendre transformations. Our results are verified numerically for an anharmonic Brownian system described in  phase space using the low-temperature quantum Fokker--Planck equations in the quantum case and the Kramers equation in the classical case, both developed for the thermodynamic system--bath model. Thus, we clarify the conditions for thermodynamics to be valid even for small systems described by Hamiltonians and establish a basis for extending thermodynamics to non-equilibrium conditions.
\end{abstract}

\maketitle

\section{Introduction}
\label{sec.intro}
Exactly 200 years have passed since Carnot published his work on the efficiency of heat engines.\cite{Carnot1824}  Thermodynamics describes macroscopic thermal phenomena in equilibrium and quasi-static processes, independently of the system dynamics. It is widely applied and has shown great success in describing thermal phenomena characterized by intensive and extensive thermodynamic properties. Massieu \cite{massieu1869} and later  Planck\cite{Planck1922} combined the total energy with temperature and entropy to derive the entropic potentials.\cite{FreeEntropyPlanes2002} Subsequently, the free energy was introduced by Gibbs and Helmholtz, completing the foundations of thermodynamics.\cite{Guggenheim1986}

Statistical mechanics, on the other hand, is a system--specific theory that describes many--body phenomena in an equilibrium state based on its microscopic statistical properties.\cite{Oono2017}
It was initiated by Boltzmann's establishment of the equality $S=k_{\rm B} \ln  {\mathcal W}$, where $S$ is the entropy, ${\mathcal W}$ is the number of possible microscopic states, and $k_{\rm B}$ is the Boltzmann constant, followed by the introduction of probability distribution functions in phase space by Gibbs and Einstein, and then by the 1922 definition of partition functions as quantum discretized eigenstates.\cite{Darwin1922} It has thus been a century since statistical mechanics took its current form.

Attempts to find a relationship between thermodynamics and statistical mechanics have been successful under limited conditions for specific systems, especially quantum systems, which are the subject of quantum thermodynamics.
\cite{PhysRevLett.97.180402,RevModPhys.81.1665,RevModPhys.83.771,Kosloff2014AnnRev,GalperinPhysRevB2015,PhysRevB.91.224303,PhysRevX.5.031044,WhitneyPhysRevB2018,NoriOtto2007,NoriDemon2009,OttoNori2020,NoriDemon2009,Hanggi2011,PhysRevBUncertenty2020,PRXQuantum.2.030202,PhysRevE.105.064124,HanggiRevModPhys2020,Kosloff_2023,PhysRevResearch.5.043274,ST21JPSJ,KT15JCP,KT16JCP} 
{However, a fundamental difference exists between the theoretical foundations of statistical physics, which is based on a first-principles description of kinetic systems from a microscopic perspective, and those of thermodynamics, which relies on  phenomenological description of thermal systems from a macroscopic viewpoint. For example,  in quantum mechanics, observables are defined as expectation values, whereas in thermodynamics, they are described by macroscopic intensive and extensive variables. 
Consequently,} a systematic theory based on a system--bath (SB) model that describes the relationship between thermodynamic potentials characterized as intensive and extensive variables has yet to be firmly established.

Recently, it has been shown that it is possible to evaluate the free energy  on the basis of an SB model from dynamically calculated work as $W \ge \Delta G$ (i.e., the minimum work principle), where $W$ is the work defined as being done from the outside to the subsystem by time-dependent external fields under an isothermal time-irreversible process, and $\Delta G$ is the change in free energy. Thus, the free energy is evaluated as $\Delta G = W^{\rm qst}$, with the system being driven quasi-statically by external fields.\cite{ST21JPSJ,ST20JCP}  Using this definition of thermodynamic potentials, the first and second laws of thermodynamics have been verified under fully quantum conditions.\cite{KT22JCP1,KT22JCP2} 

Note that the free energy defined by the partition function is ubiquitously referred to as the Helmholtz energy $\Delta F$.  
However, in the thermodynamic sense, the free energy as a function of the external field, which is an intensive variable, should be called the ``Gibbs energy,'' while the minimum work principle or the Kelvin--Planck statement is typically expressed as $W \ge \Delta F$. Since both the free energies play an essential role in the present study,  hereinafter we shall use the term ``Gibbs energy''  to refer to {what we have previously called the Helmholtz energy with reference to both our own work and that of others}.

The SB model exhibits time-irreversible dynamics toward the thermal equilibrium state of the total system owing to the infinite bath degrees of freedom, while the total equilibrium state obeys a microcanonical ensemble.\cite{Ullersma1966_1,Ullersma1966_2,CALDEIRA1983587,GRABERT1988115, Weiss2012,T06JPSJ} The effect of a heat bath is described by fluctuations and dissipation satisfying the fluctuation--dissipation theorem, and the system reaches a thermal equilibrium state in which the energy supplied by fluctuation and the energy lost by dissipation are balanced.\cite{T06JPSJ,T20JCP,T15JCP,KT13JPCB,TK89JPSJ1}  Note that in the theory of open quantum dynamics, fluctuations are essentially {non-Markovian noise expressed in terms of Matsubara frequencies, even in the case of an Ohmic spectral distribution. Therefore, it is important to adopt a non-Markovian treatment of fluctuations to describe the correct thermal equilibrium state in which the subsystem and bath are entangled (``bathentanglement'').\cite{T20JCP,T15JCP}}
Moreover, to calculate thermodynamic variables, it is necessary to accurately evaluate the heat $Q(t)$, which is the energy change in the heat bath, from the Hamiltonian dynamics.\cite{KT15JCP,KT16JCP}  
In microscopic quantum systems, it is also essential to include contributions from the SB interactions to maintain the first law of thermodynamics, which corresponds to the energy conservation law for the total system.\cite{ST20JCP,KT22JCP1,KT22JCP2}

Although the SB interaction part of the internal energy and heat, whether in quasi-static or non-equilibrium systems, is difficult to evaluate within the conventional framework of  the open quantum dynamics theory because the bath degrees of freedom have been reduced, 
we can evaluate these quantities numerically by appropriate treatment of bathentanglement, such as by using multiconfigurational time-dependent Hartree (MCTDH),\cite{VELIZHANIN2008325,TzeroWangThoss} polaron transformation,\cite{Tzero_Plenio} and other approaches.\cite{SpinBosonLettett,Tzero2003,TzeroWu,Segal2023Polaron,NoriNori2012,lambert2019modelling,OttoWiedmann_Ankerhold2021,GalperinPRL2015} 
Among these, the hierarchical equations of motion (HEOM) formalism provides a stable and versatile scheme for performing numerical simulations of a wide range of problems,\cite{TK89JPSJ1,T06JPSJ,T20JCP,IT05JPSJ,T14JCP,TW91PRA,TW92JCP,T15JCP,KT13JPCB,IT19JCTC,PhysRevB.95.064308,YiJing2022,Petruccione2023,Strunz2024}  such as for spin-lattice systems,\cite{NT18PRA} the Hornstein--Hubbard model,\cite{NT21JCP} and vibrational modes of liquid water.\cite{TT23JCP1,TT23JCP2}

Although investigations of quantum thermodynamics have progressed with the help of experimental advances, \cite{PhysRevLett.97.085901,C8TC02977F,10.1063/1.3595340,PhysRevB.84.195411} 
several issues remain unclear.
For example, extensive variables such as entropy and susceptibility, have been obtained from the free energy, since they satisfy  Legendre transformations, but how are these defined in the SB model? What is meant by the temperature derivative of the free energy evaluated from the isothermal process described by the SB model?
Are the arguments discussed in the theory of open quantum dynamics consistent with ordinary thermodynamics in the classical limit?
The answers to these questions should provide a basis for extending thermodynamics, defined by quasi-static processes, to non-equilibrium processes.

In this paper, we develop a thermodynamic theory applicable to classical and quantum systems, expressed as time-dependent thermodynamic potentials as functions of extensive and intensive variables, based on the Ullersma–Caldeira–Leggett (or Brownian) model\cite{Ullersma1966_1,Ullersma1966_2,CALDEIRA1983587,GRABERT1988115, Weiss2012}  extended for thermodynamic studies.
Our argument is based on the dimensionless minimum work principle, which is expressed using the dimensionless (or entropic) work and entropic potentials.
The results are verified by numerical simulations based on the low-temperature quantum Fokker--Planck equations (LT-QFPE)\cite{IT19JCTC}  and the Kramers equation in the classical limit.

The remainder of this paper is organized as follows: in Sec.~\ref{sec:model}, we introduce the SB Hamiltonian for thermodynamic processes. We then define non-equilibrium enthalpy and internal energies.  In Sec.~\ref{DLWHI}, we introduce dimensionless thermodynamic variables such as dimensionless entropy and polarization.
Then, in Sec.~\ref{DLMWP0}, we discuss the laws of thermodynamics within the framework of the open quantum dynamics theory. Using the quasi-static values of work and heat, we introduce the dimensionless minimum work principle and maximum entropy principle. We then express the thermodynamic potentials  in terms of intensive and extensive variables in total differential form.   In Sec.~\ref{sec:HEOM0}, to verify our results in a numerically rigorous manner, we introduce quantum and classical reduced equations of motion for thermodynamic processes and perform simulations. Finally, in Sec.~\ref{sec:conclude}, we present concluding remarks.

\section{Thermodynamic system--bath model}
\label{sec:model}

\subsection{Model Hamiltonian}
We develop a thermodynamic theory applicable to classical and quantum systems that can describe isothermal, isentropic, thermostatic, and entropic processes. To achieve this, we extend the Ullersma--Caldeira--Leggett model (or Brownian model),\cite{Ullersma1966_1,Ullersma1966_2,CALDEIRA1983587,GRABERT1988115, Weiss2012,TW91PRA,TW92JCP,T15JCP,T06JPSJ,T20JCP,KT13JPCB,IT19JCTC} which consists of a subsystem A defined in  phase space coupled to a heat bath B. By using this model, instead of the spin--boson model,\cite{SpinBosonLettett,IT05JPSJ,T14JCP} we can obtain the classical results by taking $\hbar \rightarrow 0$.  The total Hamiltonian is then expressed as
\begin{eqnarray}
\label{eq:TotalHamiltonian}
\hat{H}_{\rm tot}(t) = \hat{H}_{\rm A}(t) + \hat{H}_{\rm I+B} (t),
\end{eqnarray}
where 
\begin{eqnarray}
 \hat{H}_{\rm A}(t) =  \hat{H}_{\rm A}^0 +\hat{H}'_{\rm A} ( t ), 
\label{eq:SystemA}
\end{eqnarray}
with
\begin{eqnarray}
 \hat{H}_{\rm A}^0\equiv \frac{\hat{p}^2}{2m} + U(\hat{q})
\label{eq:SystemH}
\end{eqnarray}
being the Hamiltonian for the subsystem with mass $m$ and potential $U(\hat q)$ described by the momentum $\hat{p}$ and position $\hat{q}$, and  
\begin{eqnarray}
\label{eq:EP}
 \hat{H}'_{\rm A} ( t ) \equiv - E(t) {\mu} (\hat q)
\end{eqnarray}
is the perturbation described as ${\mu} (\hat q)$ as a function of the subsystem coordinate and the external field $E ( t )$, which is a thermodynamic-intensive variable.\cite{KT15JCP,KT16JCP} 
In a typical example, $E(t)$ denotes the laser field and ${\mu} (\hat q)$ the electric dipole moment. In laser spectroscopy, the expectation value of the dipole is an extensive variable and is observed as the polarization $P_{\rm A}(t) =\langle \mu (\hat q ) \rangle$, where $\langle \cdots \rangle$ denotes the thermal average.\cite{T06JPSJ}

Although the conventional SB model has been limited to the investigation of isothermal processes at a constant temperature, here we extend it to describe thermostatic processes in which the temperature varies with time by introducing multiple heat baths.
Thus, we consider a situation in which $N$ independent heat baths, each in the thermal equilibrium state at the inverse temperature $\beta_k\equiv 1/k_{\rm B} T_k$ are connected to or disconnected from the subsystem A using the window function $\xi_k(t)$. The bath part of the Hamiltonian is expressed as follows:
\begin{eqnarray}
\label{eq:SBHamiltonian}
\hat{H}_{\rm I+B}(t) =   \sum_{k = 1}^N  \left( \hat{H}_{\rm B}^k  + \xi_k ( t ) \hat{H}_{\rm I}^k   \right).
\end{eqnarray}
Here, the $k$th bath Hamiltonian is expressed as an ensemble of harmonic oscillators and is given by
\begin{eqnarray} 
\label{B0} 
\hat{H}_{\rm B}^k \equiv   \sum_j \left\{ \frac{ ( \hat{p}^k_j )^2 }{2 m_j} + 
\frac{1}{2} m_{j}^k (\omega_{j}^k)^2 (\hat{x}_{j}^k)^2  \right\},
\end{eqnarray}
where the momentum, position, mass, and frequency of the $j$th bath oscillator are given by $\hat{p}_{j}^k$, $\hat{x}_{j}^k$, $m_{j}^k$, and $\omega_{j}^k$, respectively.
Here, we consider the situation where the $k$th bath is always in the thermal equilibrium state $\exp(-\beta_k \hat{H}_{\rm B}^k)$ and heat is transferred to the bath only when the bath is connected to the subsystem.

The SB interaction, including the counter term, is expressed as
\begin{eqnarray}
\label{BA} 
\hat{H}_{\rm I}^k  \equiv
\sum_j \left\{ 
- A_k V(\hat{q}) c_{j}^k  \hat{x}_{j}^k  +
 \frac{ (c_{j}^k)^2 A_k^2  V^2(\hat{q})} {2 m_{j}^k(\omega_{j}^k)^2}
 \right\},
\end{eqnarray}
where  $c_{j}^k$ is the SB coupling coefficient of the $j$th bath oscillator, and $A_k$ and $V(\hat{q})$ are the coupling strength and the subsystem part of the $k$th SB interaction, respectively. 
In the study of optical spectroscopy, $V(\hat q)$ describes the effects of vibrational relaxation and dephasing.\cite{OT97PRE,TT23JCP1,TT23JCP2} The counterterm is introduced to preserve the translational symmetry of the total Hamiltonian at $U({\hat q})=0$ and to remove the undesired self-energy divergence that occurs in the Markovian case.\cite{Ullersma1966_1,Ullersma1966_2,CALDEIRA1983587,GRABERT1988115, Weiss2012,TW91PRA,TW92JCP,T15JCP,T06JPSJ,T20JCP,KT13JPCB,IT19JCTC}

The major difference when evaluating thermodynamic properties is that in the present model, the interaction term $\hat H_I^k$ is reduced with the bath, whereas in the spin-boson model,\cite{SpinBosonLettett,T14JCP} it is treated separately from the bath.\cite{KT16JCP,ST20JCP,KT22JCP1,KT22JCP2} {This makes the classical limit of the thermal equilibrium state of the time-independent subsystem  [$E(t)=0$] at $\beta_k$ identical to that of the isolated subsystem as ${\hat \rho}_{\rm A}^{\rm eq}=\exp(-\beta_k \hat H_{\rm A}^0)$, not as ${\hat \rho}_{\rm A}^{\rm eq}={\rm tr}_{\rm B}\{\exp[-\beta_k (\hat{H}_{\rm A}^0 + \hat H_{\rm I+B}^k)]\}$, while the quantum equilibrium state is still different from the isolated one owing to the bathentanglement.\cite{T20JCP,T15JCP}} 
This is a desirable feature for comparison with the conventional (classical) thermodynamic results, where the interaction is not explicitly considered.

The $k$th bath at $\beta_k$ is characterized by a spectral distribution function (SDF) defined as
\begin{eqnarray}
J^k (\omega) \equiv \sum_{j }
\frac{\hbar (c_{j}^k A_k )^2}{2m_{j}^k \omega_{j}^k} 
\delta(\omega-\omega_{j}^k).
\label{eq:J_wgeneral}
\end{eqnarray}

The window function, which we call the ``thermostatic field,'' is, for example, defined as
\begin{eqnarray}
\label{eq:xi}
 \xi_k (t)= \theta(t-t_k)\theta(t_k +\Delta t -t),
\end{eqnarray}
where  $\theta (t)$ is the step function and the time $t_k$ is defined as $t_k = t_0 + ( k - 1) \Delta t$, with  initial time $t_0$ and  duration $\Delta t$.  

For a fixed temperature $\beta_k = \beta$, $A_k ( t )\equiv A_k \xi_k(t)$ [i.e., $A_k^2( t ) =A_k^2 \xi_k(t)$] is the ``adiabatic transition field'' introduced to describe isothermal--adiabatic manipulations (e.g.,   the insertion/removal of an adiabatic wall or  the connection/disconnection of  the subsystem and  bath\cite{KT22JCP1,KT22JCP2}).  

For a fixed coupling strength $A_k = A$, the thermostatic processes can be described by utilizing $\xi_k(t)$ for multiple heat baths with different temperatures $T_k$ to set only one bath as being connected to a subsystem at a time. In this case, the bath temperature is effectively expressed as
\begin{eqnarray}
T(t)=\sum_{k = 1}^N T_k \xi_k(t),
\label{Tt}
\end{eqnarray}
or the inverse temperature as $\beta ( t )=[k_{\rm B} T ( t )]^{-1}$, except in the adiabatic case.

\subsection{Quantum fluctuation--dissipation theorem}

For the $k$th bath, if we consider the interaction coordinate of the bath modes as $\hat X^k \equiv \sum_j c_j^k \hat x_j^k$, the
subsystem A is driven by 
the external force $\hat X^k (t)$ through the interaction $-V(\hat q) \hat X^k$,
where $\hat{X}^k(t)$ 
is the Heisenberg representation of $\hat{X}^k$ for $\hat H_{\rm B}^k$.
Because each bath is a harmonic that is Gaussian in nature, the character of 
$\hat X^k (t)$ is specified by its two-time correlation functions, such as
the symmetrized and canonical correlation functions defined by\cite{TK89JPSJ1,T06JPSJ,T15JCP,T20JCP,TW91PRA,TW92JCP,T15JCP,T06JPSJ,T20JCP,KT13JPCB,IT19JCTC}  
\begin{eqnarray}
\label{eq:Xsym}
C^k(t) = \dfrac{1}{2}
\langle
\hat{X}^k(t)\hat{X}^k(0)+\hat{X}^k(0)\hat{X}^k(t)\rangle_\mathrm{B}
\end{eqnarray}
and 
\begin{align}
\label{eq:Xantisym}
\begin{split}
R^k(t) &= \frac{\beta_k}{2} \langle \hat{X}^k ; \hat{X}^k (t) \rangle_\mathrm{B}  \\
&\equiv \frac{1}{2}
\int_0^{\beta_k} {d\lambda } \, \left\langle {\hat X}^k( - {\rm i}\hbar \lambda )\hat X^k(t ) \right\rangle_\mathrm{B},
\end{split}
\end{align}
where $\langle \cdots \rangle_\mathrm{B}$ represents the thermal average of the $k$th bath degree of freedom.

The function $C^k(t)$ corresponds to fluctuation and is analogous to the classical correlation function  $X^k(t)$, whereas $R^k(t)$ corresponds to dissipation. Both are induced by the bath, and they are related through the fluctuation--dissipation theorem in  Fourier form as
$C^k[\omega] = \hbar \omega \coth ( \beta_k \hbar \omega / 2 ) R^k[\omega]$.\cite{TK89JPSJ1}  Note that in the SB system,  the fluctuation--dissipation relation is the condition to have the thermal equilibrium state, while the detailed balance condition is not satisfied {under the strong or non-Markovian SB coupling conditions}.\cite{T06JPSJ,T20JCP,T15JCP,KT13JPCB}

With the SDF, these are expressed as
\begin{eqnarray}
\label{eq:sym-exact}
C^k(t)= \int _{0}^{\infty }d\omega J^k(\omega )\coth\left(\frac{\beta_k \hbar\omega}{2}\right)\cos (\omega t) 
\end{eqnarray}
and
\begin{eqnarray}
\label{eq:rlx-exact}
R^k ( t ) = \int _{0}^{\infty }\!d\omega \,\frac{J^k(\omega )}{\hbar \omega} \cos (\omega t).
\end{eqnarray}

In any situation, by choosing the SDF in a continuous form, the bath degrees of freedom become effectively infinite. The energy eigenstates of the total system then become continuous and exhibit  time-irreversible dynamics  toward the thermal steady state, with or without time-dependent external fields, while the total system remains an isolated Hamiltonian system that conserves energy.  When $E(t)=0$, if the change in $T_k$ is small and $\Delta t$ is large, the total system enters a canonical distribution described as $\hat \rho^{\rm eq}_{\rm tot}(t_k)\approx\exp[-\beta_k (\hat{H}_{\rm A}^0 + \hat H_{\rm I+B}^k)]$, which can be verified from the steady-state solution of the HEOM and from the solution of the imaginary HEOM\cite{T15JCP} (also see Appendix~\ref{sec:MinimumWorkProof}).

In this study, we focus on the quasi-static case and assume that the correlation time of the noise is much shorter than the characteristic time scale of the system dynamics. Thus, we consider the Ohmic case described by
\begin{eqnarray}
\label{eq:ohmic-density}
J^k ( \omega ) = \frac{\hbar A_k^2 \omega}{\pi} .
\end{eqnarray}

The correlation functions in Eqs.~\eqref{eq:sym-exact} and~\eqref{eq:rlx-exact} are then evaluated as\cite{IT19JCTC} 
\begin{eqnarray}
\label{eq:sym-approx0}
\begin{split}
C^k ( t ) & \simeq \frac{2 A_k^2 }{\beta_k} \left( 1 + \sum _{l = 1}^{K} 2 \right) \delta (t) 
-\sum _{l = 1}^{K} \frac{2 A_k^2 \nu _{l}^k}{\beta_k} e^{-\nu _{l}^k | t |} 
\end{split}
\end{eqnarray}
and
\begin{eqnarray}
\label{eq:rlx-approx0}
R^k ( t ) = A_k^2 \delta (t).
\end{eqnarray}
Note that, in this Ohmic SDF, some of the physical observables, including the mean square of the momentum, $\langle p^{2}\rangle $, diverge without the cutoff $K$ because of the divergence of the first and second terms in Eq.~\eqref{eq:sym-approx0} under the infinite summation over $l$, which is often referred to as ultraviolet divergence.\cite{GRABERT1988115, Weiss2012,Ankerhold2000,Ankerhold2001}  However, because the contributions of random forces from such terms are averaged over a sufficiently short time, their effect on the dynamics of interest can be ignored by choosing a finite $K$.\cite{IT19JCTC}

We choose the coefficients $\nu _{l}^k$ and $\eta_{l}^k$ to realize the relation for finite $K$, where the first term on the right-hand side of Eq.~\eqref{eq:sym-approx0} is the classical contribution from the temperature, and the remaining terms are the quantum low-temperature (QLT) corrections.\cite{IT19JCTC} It should be noted that the fluctuation term is always non-Markovian because of the quantum nature of the noise; it can be regarded as Markovian only in the high-temperature limit, $\beta_k \hbar \omega _{\mathrm{0}}\ll 1$, in which the heat bath exhibits a classical behavior.\cite{T06JPSJ,Weiss2012} This is an important conclusion obtained from the quantum fluctuation--dissipation theorem. That is, negative non-Markov terms always appear in the Ohmic case unless  a time coarse-grained Markov assumption is adopted, which is often unphysical as a description of quantum thermodynamics.\cite{T06JPSJ,T20JCP,T15JCP,KT13JPCB}

\subsection{Non-equilibrium enthalpy and internal energy}
\label{sec:Internal}

In a typical thermodynamic theory, a system is characterized by thermodynamic potentials described in terms of intensive variables such as electric field $E$ and temperature $T$, and extensive variables, such as polarization $P$ and entropy $S$.  To construct a quantum thermodynamic theory similar to thermodynamics,  isothermal and isentropic (d$E$ and d$P$) processes as well as thermostatic and entropic (d$T$ and d$S$) processes must be investigated. However, most investigations, including traditional thermodynamics, to date have been limited to isothermal processes in which only the external field corresponding to the intensive variable is manipulated.

Although various quantum thermodynamic studies have been conducted, it has been difficult to derive thermodynamic laws involving the total derivative of temperature (d$T$) owing to the difficulty of including thermostatic processes. Moreover, these studies have not introduced extensive variables, such as $P$, as observables in open quantum systems, nor have they provided their Legendre transformations, which play an essential role in thermodynamics.

Here, we show that we can construct a complete description of a thermodynamic theory from work in isothermal processes ($T_{\rm fix}$) and heat in constant external field (or thermostatic) processes ($E_{\rm fix}$). For this purpose, we introduce the expectation values of energy for each component of the Hamiltonian.
% improve readability
{From the Hamiltonian [Eqs.~\eqref{eq:SystemA}--\eqref{eq:EP}], the total energy of the subsystem}  is expressed as
\begin{eqnarray}
\label{eq:WA}
H_{\rm A} (t) = U_{\rm A} ( t ) +  H' ( t ) ,
\end{eqnarray}
where
\begin{eqnarray}
\label{eq:WA0}
U_{\rm A} (t) \equiv \mathrm{tr}_{\rm A} \{  \hat{H}_{\rm A}^0 \hat{\rho}_{\rm A} ( t )\}
\end{eqnarray}
is the self-energy of the subsystem, which is considered ``internal energy'' in a quasi-static case, and
\begin{eqnarray}
\label{eq:WAEP}
H'(t) = - E(t) P_{\rm A}(t)
\end{eqnarray}
is the interaction energy described with the optical polarization defined as
\begin{eqnarray}
\label{eq:P}
P_{\rm A}(t) \equiv  \mathrm{tr}_{\rm A}\left\{ {\mu} (\hat q) \hat{\rho}_{\rm A} ( t ) \right\},
\end{eqnarray}
where $\hat{\rho}_{\rm A}( t )=\mathrm{tr}_{\rm B}\{\hat{\rho}_{\rm tot} ( t ) \}$ is the reduced density operator for the total density operator $\hat{\rho}_{\rm tot} ( t )$.  

While $E(t)$ is an intensive variable, $P_{\rm A}(t)$ is an extensive variable. This is because for distinguishable $M$ subsystems without mutual interaction, the total density operator is expressed as $\hat{\rho}_{\rm A}^{( M )} = \hat{\rho}_{\rm A} \otimes \cdots \otimes \hat{\rho}_{\rm A}$, and thus $P_{\rm A} (t)$ becomes proportional to the size of the system.  Moreover, because the evaluation of $P_{\rm A}(t)$ requires a statistical average of the distribution function $\hat{\rho}_{\rm A} ( t )$ even for a single-particle system, we can evaluate it as a thermal variable.

% improve readability
Because $E(t)$ and $P_{\rm A}(t)$ are conjugate to each other, the relation 
\begin{eqnarray}
\label{eq:LegendreH-U}
H_{\rm A} ( t ) = U_{\rm A} ( t ) - E ( t ) P_{\rm A} ( t )
\end{eqnarray}
is regarded as a time-dependent Legendre transformation between non--equilibrium enthalpy $H_{\rm A} ( t )$ and internal energy $U_{\rm A} ( t )$.

{The total non-equilibrium enthalpy is now expressed as
\begin{eqnarray}
\label{eq:Utot}
H_{\rm tot}(t) = H_{\rm A} (t) + \sum_{k = 1}^N H_{\rm I + B}^k ( t ),
\end{eqnarray}
where $H_{\rm I + B}^k ( t )$ is the enthalpy of the $k$th bath, defined as
\begin{eqnarray}
H_{\rm I + B}^k ( t ) = \mathrm{tr}_{\rm tot} \left\{ \hat{H}_{\rm I + B}^k ( t )
\hat{\rho}_{\rm tot} ( t ) \right\} .
\end{eqnarray}
The bath part of the energy (non-equilibrium bath enthalpy) is then expressed as}
\begin{eqnarray}
\label{eq:UB}
H_{\rm I + B} ( t )  \equiv \sum_{k = 1}^N H_{\rm I + B}^k ( t ) .
\end{eqnarray}

\section{Dimensionless thermodynamic variables}
\label{DLWHI}

Although the minimum work principle for an intensive variable, expressed as $W_{\rm A}^{int} (t) \ge \Delta G_{\rm A}^{\rm qst}$, provides the condition for determining the thermodynamic potential from work,\cite{ST20JCP} it does not do so for heat, which is important in determining entropy. We find that this difficulty can be overcome by introducing the dimensionless enthalpy of the subsystem, defined as $\tilde{H}_{\rm A}(t)\equiv \beta(t)H_{\rm A}(t)$.  We use this to define the changes in dimensionless {(or entropic)} ``intensive work'' and ``extensive heat'' in time corresponding to power and heat flow, respectively,  as follows:
\begin{eqnarray}
\label{DefWint}
\frac{d \tilde{W}_{\rm A}^{int} ( t )}{d t}
\equiv \mathrm{tr}_{\rm A} \left\{ \frac{\partial}{\partial t}  \left[ \beta ( t ) \hat{H}_{\rm A} ( t )  \right]
\hat{\rho}_{\rm A} ( t ) \right\} 
\end{eqnarray}
and
\begin{eqnarray}
\label{DefQ}
\frac{d \tilde{Q}_{\rm A}^{ext} ( t )}{d t} \equiv  \mathrm{tr}_{\rm A} \left\{ \left[ \beta ( t ) \hat{H}_{\rm A} ( t )  \right]
\frac{\partial \hat{\rho}_{\rm A} ( t )}{\partial t} \right\} . 
\end{eqnarray}
These are interrelated by the following time-dependent Legendre transformation:
\begin{eqnarray}
\label{eq:LegendreWint-Q}
\begin{split}
\frac{d \tilde{W}^{int}_{\rm A} ( t )}{d t} &= - \frac{d \tilde{Q}_{\rm A}^{ext}  ( t )}{d t}
+ \frac{d}{d t} \left[ \beta ( t ) H_{\rm A} ( t ) \right] .
\end{split}
\end{eqnarray}
They are then evaluated as
\begin{eqnarray}
\label{ITwork}
\frac{d \tilde{W}^{int}_{\rm A} ( t )}{d t} =  H_{\rm A} ( t ) \frac{d \beta ( t )}{d t} - \tilde{P} _{\rm A} ( t ) \frac{d E ( t )}{d t}
\end{eqnarray}
and
\begin{eqnarray}
\label{eq:QA2}
\frac{d \tilde{Q}_{\rm A}^{ext}  ( t )}{d t}
= \beta ( t ) \frac{d H_{\rm A} ( t )}{d t} + \tilde{P} _{\rm A}( t ) \frac{d E ( t )}{d t},
\end{eqnarray}
where $\tilde{P} _{\rm A} ( t )$ is the dimensionless polarization,  defined as
\begin{eqnarray}
\tilde{P}_{\rm A} ( t ) = \beta ( t ) P _{\rm A}( t ) .
\end{eqnarray}
In quantum thermodynamics, work has been defined using the change in an intensive variable as $P_{\rm A} ( t ) d E ( t )$. 
In typical thermodynamics, however, work can also be defined as the change in an extensive variable as $-E ( t ) d P_{\rm A} ( t )$. To treat  work defined as such, we further introduce the dimensionless (or entropic) ``extensive work'' using a time-dependent Legendre transformation of $\tilde{W}^{int}_{\rm A} ( t )$ as follows:
\begin{eqnarray}
\label{eq:DefWext}
\begin{split}
\frac{d \tilde{W}_{\rm A}^{ext} ( t )}{d t} &= \frac{d \tilde{W}^{int}_{\rm A} ( t )}{d t}
+ \frac{d}{d t} \left[ \tilde{P}_{\rm A} ( t ) E ( t ) \right] \\
&
= H_{\rm A} ( t ) \frac{d \beta ( t )}{d t}
+ E ( t ) \frac{d \tilde{P}_{\rm A} ( t )}{d t} .
\end{split}
\end{eqnarray}

The fundamental difference between the present treatment and the conventional treatment of thermodynamics is that the total energy, including the heat bath, is explicitly described by the Hamiltonian. Thus, we consider the dimensionless total enthalpy expressed as
\begin{eqnarray}
\label{eq:DefHtot}
\tilde{H}_{\rm tot} ( t )  
= \mathrm{tr}_{\rm tot} \left\{ \hat{\tilde{H}}_{\rm tot} ( t ) \hat{\rho}_{\rm tot} ( t ) \right\} ,
\end{eqnarray}
where $\hat{\tilde{H}}_{\rm tot} ( t )$ is the dimensionless total Hamiltonian defined as
\begin{eqnarray}
\label{eq:DefTHtot}
\hat{\tilde{H}}_{\rm tot} ( t ) \equiv \beta ( t ) \hat{H}_{\rm A} ( t )
+ \sum_{k = 1}^N \beta_k \hat{H}_{\rm I + B}^k ( t ) .
\end{eqnarray}
The relationship between dimensionless (or entropic) total work and heat is then expressed as
\begin{eqnarray}
\label{eq:LTSB} 
\frac{d \tilde{W}_{\rm tot} ^{int}  ( t )}{d t} = - \frac{d \tilde{Q}_{\rm tot}^{ext}  (t) }{d t} 
+  \frac{d }{d t} \tilde{H}_{\rm tot} ( t ),
\end{eqnarray}
where
\begin{eqnarray}
\label{eq:therQB1}
 \quad \frac{d \tilde{W}_{\rm tot} ^{int}  ( t ) }{d t} 
= \mathrm{tr}_{\rm tot} \left\{ \frac{\partial \hat{\tilde{H}}_{\rm tot} ( t )}{\partial t} 
\hat{\rho}_{\rm tot} ( t ) \right\}
\end{eqnarray}
and
\begin{eqnarray}
\label{eq:entQB2}
\frac{d \tilde{Q}_{\rm tot}^{ext}  ( t ) }{d t} = 0 .
\end{eqnarray}
Here, to obtain Eq.~\eqref{eq:entQB2}, we consider the case in which the subsystem is connected to only one bath at the same time and use the following identity:
\begin{eqnarray}
\label{eq:totalzero}
\mathrm{tr}_{\rm tot} \left\{ \hat H_{\rm tot}(t) {\frac{\partial \hat{\rho}_{\rm tot} ( t )}{\partial t}} \right\} = 0 .
\end{eqnarray}
From Eq.~\eqref{eq:LTSB}, we have 
\begin{eqnarray}
\label{eq:LTSB2} 
\frac{d \tilde{W}_{\rm tot} ^{int}  ( t )}{d t} = \frac{d \tilde{H}_{\rm tot}(t)}{d t} .
\end{eqnarray}
Thus, the thermal current between the subsystem and bath is conserved as
\begin{eqnarray}
\label{eq:extcancel2}
\frac{d \tilde{Q}_{\rm A}^{ext}  ( t )}{d t} = - \frac{d \tilde{Q}_{\rm I + B}^{ext}  ( t )}{d t},
\end{eqnarray}
where 
\begin{eqnarray}
\label{eq:entQB3}
\frac{d \tilde{Q}_{\rm I + B}^{ext}  ( t ) }{d t} = \sum_{k = 1}^N
{\mathrm{tr}_{\rm tot} \left\{ \beta_k \hat{H}_{\rm I + B}^k ( t )
\frac{\partial \hat{\rho}_{\rm tot} ( t )}{\partial t} \right\} }.
\end{eqnarray}
We then identify the time derivative of the dimensionless bath entropy, i.e., {$\tilde{S}_{\rm I + B}  ( t ) = {S}_{\rm I + B}  ( t ) /k_{\rm B}$,}  with the dimensionless bath heat current as follows:
\begin{eqnarray}
\label{eq:DefBathEntropy}
{\frac{d \tilde{S}_{\rm I + B} ( t )}{d t} = \frac{d \tilde{Q}_{\rm I + B}^{ext}  ( t )}{d t}.}
\end{eqnarray}

\section{The laws of thermodynamics}
\label{DLMWP0}

Quantum mechanics is a first-principles theory, and its logical structure is reasonably simple. By contrast, thermodynamics is a phenomenological theory developed on the basis of several principles and statements. In this section, we deduce thermodynamics from quantum mechanics using observables, defined as quantum mechanical expectation values, to clarify the central dogma of thermodynamics. 

\subsection{First to third laws of thermodynamics}
\label{LofTD}

\subsubsection{First law: Energy conservation law}

The key principle for energy in quantum mechanics is the energy conservation law for work, whereas in thermodynamics, energy is described not only as work but also as heat. In fact, by adding Eqs.~\eqref{eq:QA2} and~\eqref{eq:DefWext}, we have the first law of thermodynamics expressed as
\begin{eqnarray}
\label{eq:totalwork}
\frac{d \tilde U_{\rm A} ( t )}{d t}=\frac{d \tilde{W}_{\rm A}^{ext} ( t )}{d t}+\frac{d \tilde{Q}_{\rm A}^{ext}  ( t )}{d t},
\end{eqnarray}
where
\begin{eqnarray}
\label{StateVar}
\tilde U_{\rm A} ( t ) =\beta ( t ) H_{\rm A} ( t ) + E ( t ) \tilde{P}_{\rm A}  ( t ) 
\end{eqnarray}
is the dimensionless non-equilibrium internal energy also expressed as $\tilde U_{\rm A} ( t ) =\beta(t) U_{\rm A} ( t )$. The above-mentioned equality is the consequence of the energy conservation law expressed as the time derivative of Eq.~\eqref{eq:Utot} as follows:
\begin{eqnarray}
\label{Total1stLaw}
\frac{d H_{\rm tot} ( t )}{d t} = \frac{d H_{\rm A} ( t )}{d t} + \frac{d H_{\rm I+B} ( t )}{d t}.
\end{eqnarray}
Then, using Eqs.~\eqref{eq:totalzero} and~\eqref{eq:extcancel2}, we obtain Eq.~\eqref{eq:totalwork}.

\subsubsection{Second law: Increasing internal energy under time-irreversible process}

As an extension of the minimum work principle for work as being done from the outside to the subsystem, we consider ``the dimensionless {(or entropic)} minimum work principle'' for the total system from one equilibrium state to another, expressed as
\begin{eqnarray}
\label{eq:TheMinimumPrinciple1}
\tilde{W}_{\rm tot}^{int} \geq \left( \tilde{W}^{int}_{\rm tot} \right)^{\rm qst} ,
\end{eqnarray}
where $\tilde{W}_{\rm tot}^{int}$ is defined by Eq.~\eqref{eq:therQB1} and $( \tilde{W}^{int}_{\rm tot} )^{\rm qst}$ represents a transition that occurs quasi-statically. 
The proof of the above-mentioned inequality is given in Appendixes~\ref{sec:MinimumWorkProof} and~\ref{sec:qstProof}. Unlike the existing minimum work principle, this definition can treat a thermostatic process.
From Eq.~\eqref{eq:LTSB2}, the above-mentioned inequality can also be expressed as
\begin{eqnarray}
\label{eq:TheMinimumPrinciple2}
\Delta \tilde{H}_{\rm tot} \geq \Delta \tilde{H}_{\rm tot}^{\rm qst},
\end{eqnarray}
which corresponds to ``the law of increasing enthalpy'' under a time-irreversible process for a closed system. 

When the coupling strength is fixed and the contribution of work from the SB interaction is incorporated into the system, which is the case in the present Brownian-based model, we have the equality
\begin{eqnarray}
\frac{d \tilde{W}_{\rm A}^{int} ( t )}{d t} = \frac{d \tilde{W}_{\rm tot}^{int} ( t )}{d t} ,
\end{eqnarray}
and Eq.~\eqref{eq:TheMinimumPrinciple1} reduces to
\begin{eqnarray}
\label{eq:MinimumPrinciple}
\tilde{W}_{\rm A}^{int} \geq \left( \tilde{W}^{int}_{\rm A} \right)^{\rm qst}.
\end{eqnarray}
This relation is an extension to thermostatic processes of the minimum work principle for isothermal processes, which states that thermodynamic weight is maximized in a quasi-static equilibrium state.
Because not only the intensive variables $\beta ( t )$ and $E ( t )$ but also the extensive variables $\tilde{P}_{\rm A} ( t )$ and $H_{\rm A} ( t )$ may change independently during a non-equilibrium transition between the two equilibrium states, we can obtain the relationship for heat with the use of Eq.~\eqref{eq:LegendreWint-Q} as
\begin{eqnarray}
\label{eq:MaximumDLHeat}
\tilde{Q}_{\rm A}^{ext}  \leq \left( \tilde{Q}_{\rm A}^{ext} \right)^{\rm qst} ,
\end{eqnarray}
which is equivalent to ``the principle of maximum entropy.'' This is because, for the transition from equilibrium state 1 at time $t_1$ to equilibrium state 2 at $t_2$, we have 
\begin{eqnarray}
\label{eq:MaximumS}
\int^{t_2}_{t_1} \frac{1}{T ( t' )} \frac{d Q_{\rm A} ( t' )}{d t'} d t'  \leq  \Delta S_{\rm A}^{\rm qst} ,
\end{eqnarray}
where
\begin{eqnarray}
\label{eq:DefQA}
{\frac{d Q_{\rm A} ( t )}{d t} = \mathrm{tr}_{\rm A} \left\{ \hat{H}_{\rm A} ( t ) 
\frac{\partial \hat{\rho}_{\rm A} ( t )}{\partial t} \right\} }
\end{eqnarray}
is the system heat current,\cite{KT16JCP} which satisfies $\beta ( t ) d Q_{\rm A} ( t )/dt  = d \tilde{Q}_{\rm A}^{ext} ( t )/dt$.

From Eqs.~\eqref{eq:extcancel2} and~\eqref{eq:MaximumS}, the law of total entropy production is expressed as
\begin{eqnarray}
\Delta S_{\rm tot} \geq 0 ,
\end{eqnarray}
where $\Delta S_{\rm tot}$ is a total entropy difference between the equilibrium states at time $t_1$ and $t_2$, defined as
\begin{eqnarray}
\Delta S_{\rm tot} =
{\Delta S_{\rm A}^{\rm qst}} + \int^{t_2}_{t_1} \frac{d S_{\rm I + B} ( t )}{d t} d t .
\end{eqnarray}
Here, $S_{\rm I + B} ( t ) = k_{\rm B} \tilde{S}_{\rm I + B} ( t )$ is the bath entropy [see Eq.~\eqref{eq:DefBathEntropy}].

We can also obtain the inequality for dimensionless (or entropic) extensive work 
using the Legendre transformation~\eqref{eq:DefWext} for Eq.~\eqref{eq:MinimumPrinciple} as
\begin{eqnarray}
\label{eq:MinimumExtWork}
\tilde{W}_{\rm A}^{ext} \geq \left( \tilde{W}_{\rm A}^{ext} \right)^{\rm qst}.
\end{eqnarray}

\subsubsection{Third law: Zero-temperature limit of the SB model}

In the framework of open quantum dynamics theory, the third law of thermodynamics should describe the state of the subsystem in the limit of zero temperature, but anomalous behavior  has been found in the spin-boson system at zero temperature.\cite{SpinBosonLettett,Tzero2003,TzeroWangThoss,Tzero_Plenio,TzeroWu} Thus, entropy may not be zero even at zero temperature because of the  essential role played by  bathentanglement: the entropy may diverge as $\beta \rightarrow \infty$ as the classical approximation ($\beta \hbar \omega_{\rm A} \ll 1$, where $\omega_{\rm A}$ is the characteristic frequency of the subsystem) collapses in the low-temperature regime. This implies that some statements of the third law of thermodynamics, such as ``the entropy of a system at absolute zero is a well-defined constant,'' do not hold for quantum systems described by an SB model, whereas Nernst's statement ``it is impossible for any procedure to reach $T = 0$ in a finite number of steps,'' 
which is equivalent to the assertion that ``thermodynamic states with $T=0$ do not exist'' seems reasonable---although theoretically anything is possible.

\subsection{Thermodynamic potentials}

\subsubsection{Massieu--Planck potentials: Entropic state representation}
\label{Massieu}

Quasi-static dimensionless (or entropic) work and heat are related to the (dimensionless) Massieu--Planck potentials, which are entropic thermodynamic potentials defined as  Legendre transforms of entropy.\cite{Guggenheim1986,massieu1869,Planck1922,FreeEntropyPlanes2002} Two of these functions are the Massieu and Planck potentials, introduced as  dimensionless forms of the Helmholtz and Gibbs energies and defined as $\Phi_{\rm A}  \equiv \beta F_{\rm A}$ and $\Xi_{\rm A}  \equiv \beta G_{\rm A}$, respectively. The other two are entropy--derived potentials: the dimensionless entropy $\Pi_{\rm A} $ and its conjugate entropy  $\Theta_{\rm A} $.
Because these two do not seem to have previously been given names, here we refer to $\Pi_{\rm A}$ as the Clausius entropy (C-entropy) after Clausius, who first introduced entropy,\cite{Oono2017} and to $\Theta_{\rm A}$, associated with the partition function that manifestly includes an intensive variable, as the Boltzmann entropy (B-entropy). In the present case, we can introduce the Massieu and Planck potentials and the B-entropy as $\Delta \Phi_{\rm A}^{\rm qst} = - ( W_{\rm A}^{ext} )^{\rm qst}$,  $\Delta \Xi_{\rm A}^{\rm qst} = - ( W_{\rm A}^{int} )^{\rm qst},$ and $\Delta \Theta_{\rm A}^{\rm qst} = ( \tilde{Q}_{\rm A}^{ext} )^{\rm qst}$, respectively. 

Thus, from Eqs.~\eqref{eq:MinimumPrinciple} and~\eqref{eq:MinimumExtWork}, we obtain the dimensionless (or entropic) minimum work principle for the intensive and expensive work defined as being done from the outside as
\begin{eqnarray}
\label{eq:MinimumPrinciple0}
\tilde{W}_{\rm A}^{int} \geq - \Delta \Xi_{\rm A}^{\rm qst} 
\end{eqnarray}
and
\begin{eqnarray}
\label{eq:MinimumExtWork0}
\tilde{W}_{\rm A}^{ext} \geq -\Delta \Phi_{\rm A}^{\rm qst}.
\end{eqnarray}
The latter inequality is a generalization of the Kelvin--Planck statement, often used as a definition of the second law of thermodynamics, for isothermal processes. From Eq.~\eqref{eq:MaximumDLHeat}, we obtain
\begin{eqnarray}
\label{eq:MaximumDLHeat0}
\tilde{Q}_{\rm A}^{ext}  \leq \Delta \Theta_{\rm A}^{\rm qst},
\end{eqnarray}
which we call ``the principle of maximum dimensionless heat generation'' and which is equivalent to  ``the principle of maximum B-entropy.''

\begin{table*}[!t]
\caption{\label{table:DLPotential}  Total differential expressions for the quasi-static (qst.) entropic potentials as functions of the intensive variables $\beta^{\rm qst}(t)$ and  $E^{\rm qst}(t)$ and the extensive variables $H_{\rm A}^{\rm qst}(t)$ and  $\tilde{P}_{\rm A}^{\rm qst}(t)$. Entropy has two definitions, depending on whether the work variable is intensive or extensive.
Of these dimensionless entropies, the commonly used one, which we call Clausius entropy (C-entropy) and involves only extensive variables, is expressed as $\Pi_{\rm A}^{\rm qst}[H_{\rm A}^{\rm qst}, P_{\rm A}^{\rm qst} ]$, 
whereas the less widely used one, which we call Boltzmann entropy (B-entropy),  is expressed as $\Theta_{\rm A}^{\rm qst} [ H_{\rm A}^{\rm qst}, E^{\rm qst} ]$. The potentials are related by the Legendre transformations as shown.}
\begin{ruledtabular}
\begin{tabular}{llcc}
Qst. Potential & Differential form & Natural var. & Legendre transformation 
\\
\hline
Massieu & 
$d \Phi_{\rm A}^{\rm qst}  = - H_{\rm A}^{\rm qst} d \beta^{\rm qst}  - E^{\rm qst} d \tilde{P}_{\rm A}^{\rm qst}$ 
& $\beta^{\rm qst} , \tilde{P}_{\rm A}^{\rm qst}$ & $\cdots$ 
\\
Planck  & 
$d \Xi_{\rm A}^{\rm qst}  = - H_{\rm A}^{\rm qst} d \beta^{\rm qst} + \tilde{P}_{\rm A}^{\rm qst} d E^{\rm qst}$
& $\beta^{\rm qst} , E^{\rm qst}$ & $\Xi_{\rm A}^{\rm qst} = \Phi_{\rm A}^{\rm qst} + E^{\rm qst} \tilde{P}_{\rm A}^{\rm qst}$
\\
B-Entropy 
& $d \Theta_{\rm A} ^{\rm qst} = \beta^{\rm qst} d H_{\rm A}^{\rm qst} + \tilde{P}_{\rm A}^{\rm qst} d E^{\rm qst}$
& $H_{\rm A}^{\rm qst} , E^{\rm qst}$
& $\Theta_{\rm A} ^{\rm qst} = \Xi_{\rm A} ^{\rm qst} + \beta^{\rm qst} H^{\rm qst}_{\rm A} $
\\
C-Entropy 
& $d \Pi_{\rm A}^{\rm qst}  = \beta^{\rm qst} d H_{\rm A}^{\rm qst} - E^{\rm qst} d \tilde{P}_{\rm A}^{\rm qst}$
& $H_{\rm A}^{\rm qst} , \tilde{P}_{\rm A}^{\rm qst}$
& $\Pi_{\rm A}^{\rm qst}  =  \Phi_{\rm A}^{\rm qst} + \beta^{\rm qst} H^{\rm qst}_{\rm A} $
\end{tabular}
\end{ruledtabular}
\end{table*}

From Eqs.~\eqref{eq:LegendreWint-Q} and~\eqref{eq:DefWext}, we find that these entropic potentials are related by the following Legendre transformations:
\begin{eqnarray}
\label{eq:LegendreP-M}
\Phi_{\rm A}^{\rm qst} ( t ) = \Xi_{\rm A}^{\rm qst} ( t ) - \tilde{P}_{\rm A}^{\rm qst}(t) E^{\rm qst} ( t ) 
\end{eqnarray}
and
\begin{eqnarray}
\label{eq:LegendreP-E}
\Theta_{\rm A}^{\rm qst} ( t ) = \Xi_{\rm A}^{\rm qst} ( t ) 
+ \beta^{\rm qst} ( t ) H_{\rm A}^{\rm qst} ( t ) ,
\end{eqnarray}
respectively, where the quasi-equilibrium values of the potentials are evaluated from Eqs.~\eqref{eq:MinimumPrinciple},~\eqref{eq:MaximumDLHeat}, and~\eqref{eq:MinimumExtWork}.  From these, the C-entropy can be expressed as 
\begin{eqnarray}
\label{eq:LegendreP-S}
\Pi_{\rm A}^{\rm qst} ( t ) = \Phi_{\rm A}^{\rm qst} ( t ) 
+ \beta^{\rm qst} ( t ) H_{\rm A}^{\rm qst} ( t ).
\end{eqnarray}

The Planck potential is convex for the inverse temperature $\beta^{\rm qst} ( t )$ and the external field $E^{\rm qst} ( t )$ (see Appendix~\ref{sec:Convexity}). Thus, the natural variables of the Planck potential are $\beta^{\rm qst} ( t )$ and $E^{\rm qst} ( t )$, described as $\Xi_{\rm A}^{\rm qst} [ \beta^{\rm qst} ( t ) ,\,  E^{\rm qst} ( t ) ]$. From Eqs.~\eqref{eq:LegendreP-M}--\eqref{eq:LegendreP-S}, we can express the Massieu potential and the dimensionless entropies as $\Phi_{\rm A}^{\rm qst} [ \beta^{\rm qst} ( t ), \, \tilde{P}_{\rm A}^{\rm qst} ( t ) ]$, $\Theta_{\rm A}^{\rm qst} [ H_{\rm A}^{\rm qst} ( t ), \, E^{\rm qst} ( t ) ]$, and  $\Pi_{\rm A}^{\rm qst} [ H_{\rm A}^{\rm qst} ( t ),\, \tilde{P}_{\rm A}^{\rm qst} ( t ) ]$. The various relationships between the entropic potentials and intensive and extensive variables can be derived for these potentials. \cite{Guggenheim1986,FreeEntropyPlanes2002} It should be noted that although we chose enthalpy as the natural variable, it is possible to choose internal energy for the description of the dimensionless thermodynamic potentials (see Appendix~\ref{sec:InternalRep}).

We summarize the dimensionless thermodynamic (or entropic)  potentials as functions of the natural variables in total differential form in Table~\ref{table:DLPotential}.

\subsubsection{Helmholtz--Gibbs potentials: Energy state representation}
\label{Thermopotential}

From dimensionless thermodynamic variables, potentials are naturally expressed in the entropic representation, while 
the energy state representation of potentials is commonly used in thermodynamics.  For convenience in numerical simulations and to facilitate extension to non-equilibrium states,\cite{ST20JCP, KT22JCP1, KT22JCP2} we introduce these potentials by evaluating them for the isothermal case and constant-external field   (thermostatic) cases.

For  fixed $\beta ( t ) = \beta$ (isothermal case), the inequalities~\eqref{eq:MinimumPrinciple} and~\eqref{eq:MinimumExtWork} for $F_{\rm A}^{\rm qst} ( t ) = - ( \tilde{W}_{\rm A}^{ext})^{\rm qst} / \beta$ and  $G_{\rm A}^{\rm qst} ( t ) = - ( \tilde{W}_{\rm A}^{int})^{\rm qst} / \beta$ reduce to
\begin{eqnarray}
\label{eq:ExtKelvinPlanck}
W_{\rm A}^{ext} \geq \Delta F_{\rm A}^{\rm qst}
%\left(  W_{\rm A}^{ext} \right)^{\rm qst}
\end{eqnarray}
and
\begin{eqnarray}
\label{eq:IntKelvinPlanck}
W_{\rm A}^{int} \geq \Delta G_{\rm A}^{\rm qst},
%\left( W_{\rm A}^{int} \right)^{\rm qst},
\end{eqnarray}
where we define the extensive work $W_{\rm A}^{ext}$ and intensive work $W_{\rm A}^{int}$ by
\begin{eqnarray}
\label{eq:DefWext2}
\frac{d W_{\rm A}^{ext} ( t )}{d t}
%= \frac{1}{\beta} \frac{d \tilde{W}_{\rm A}^{ext} ( t )}{d t}
= - \mathrm{tr}_{\rm A} \left\{ \hat{H}'_{\rm A} ( t ) \frac{\partial \hat{\rho}_{\rm A} ( t )}{\partial t} \right\}
\end{eqnarray}
and
\begin{eqnarray}
\label{eq:DefWint2}
\frac{d W_{\rm A}^{int} ( t )}{d t} % = \frac{1}{\beta} \frac{d \tilde{W}_{\rm A}^{int} ( t )}{d t}
= \mathrm{tr}_{\rm A} \left\{ \frac{\partial \hat{H}'_{\rm A} ( t )}{\partial t} \hat{\rho}_{\rm A} ( t ) \right\}.
\end{eqnarray}
We can then  evaluate the two free energies as $\Delta F_{\rm A}^{\rm qst}=(  W_{\rm A}^{ext})^{\rm qst}$ and $\Delta G_{\rm A}^{\rm qst}=( W_{\rm A}^{int})^{\rm qst}$. The inequality~\eqref{eq:ExtKelvinPlanck} corresponds to the Kelvin--Planck statement. From Eq.~\eqref{eq:DefWext}, these intensive and extensive works satisfy the time-dependent Legendre transformation, expressed as
\begin{eqnarray}
\frac{d W_{\rm A}^{ext} ( t )}{d t} = \frac{d W_{\rm A}^{int} ( t )}{d t}
+ P_{\rm A} ( t ) E ( t ) .
\end{eqnarray}

While the Planck potential is convex for the external field $E^{\rm qst} ( t )$, the Gibbs energy is concave for $E^{\rm qst} ( t )$,  because $G_{\rm A}^{\rm qst} ( t ) = -\Xi_{\rm A}^{\rm qst} (t)/\beta$. From Eq.~\eqref{eq:LegendreP-M}, the Helmholtz and Gibbs energies satisfy the Legendre transformation expressed as
\begin{eqnarray}
\label{eq:LegendreG-F}
F_{\rm A}^{\rm qst} ( t ) = G_{\rm A}^{\rm qst} ( t ) + P_{\rm A}^{\rm qst} ( t ) E^{\rm qst} ( t ) .
\end{eqnarray}

\begin{table*}[!t]
\caption{\label{table:Potential} Total differential expressions for the quasi-static (qst.)  thermodynamic potentials as  functions of the intensive variables $T^{\rm qst}(t)$ and  $E^{\rm qst}(t)$ and the extensive variables $S_{\rm A}^{\rm qst}(t)$ and  ${P}_{\rm A}^{\rm qst}(t)$, which are related through  Legendre transformations.}
\begin{ruledtabular}
\begin{tabular}{llcc}
Qst.  potential & Differential form & Natural var. & Legendre transformation 
\\
\hline
Helmholtz 
& $d F_{\rm A} ^{\rm qst} = - S_{\rm A} ^{\rm qst} d T^{\rm qst} + E^{\rm qst} d P_{\rm A}^{\rm qst}$
& $T^{\rm qst} , P_{\rm A}^{\rm qst}$
& $\cdots$ 
\\
Gibbs  
& $d G_{\rm A} ^{\rm qst} = - S_{\rm A} ^{\rm qst} d T^{\rm qst} - P_{\rm A}^{\rm qst} d E^{\rm qst}$
& $T^{\rm qst} , E^{\rm qst}$ 
& $G_{\rm A}^{\rm qst} = F_{\rm A}^{\rm qst} - E^{\rm qst} P_{\rm A}^{\rm qst}$ 
\\
Internal 
& $d U_{\rm A}^{\rm qst} = T^{\rm qst} d S_{\rm A}^{\rm qst} + E^{\rm qst} d P_{\rm A}^{\rm qst}$
& $S_{\rm A}^{\rm qst} , P_{\rm A}^{\rm qst}$
& $U_{\rm A}^{\rm qst} =  F_{\rm A}^{\rm qst} + T^{\rm qst} S_{\rm A}^{\rm qst}$
\\
Enthalpy
& $d H_{\rm A}^{\rm qst} = T^{\rm qst} d S_{\rm A}^{\rm qst} - P_{\rm A}^{\rm qst} d E^{\rm qst}$
& $S_{\rm A}^{\rm qst} , E^{\rm qst}$
& $H^{\rm qst}_{\rm A} = G_{\rm A} ^{\rm qst} + T^{\rm qst} S_{\rm A}^{\rm qst}$
\end{tabular}
\end{ruledtabular}
\end{table*}

When we fix the external field $E ( t ) = E_{\rm fix}$, the inequality for the dimensionless entropy reduces to 
\begin{eqnarray}
\int^{t_2}_{t_1} \beta ( t ) \frac{d H_{\rm A} ( t )}{d t} d t \leq \Delta \Theta_{\rm A}^{\rm qst} ,
\end{eqnarray}
where the system is in equilibrium at times $t_1$ and $t_2$ and we have introduced the time-integral form so that $H_{\rm A}(t)$ 
can be treated even when it has a singular point as a function of $t$. From Appendix~\ref{sec:Convexity}, we can prove that the dimensionless entropy is concave as a function of enthalpy, expressed as
\begin{eqnarray}
\label{eq:Concavity}
\frac{\partial^2 \Theta_{\rm A}^{\rm qst} ( t )}{\partial \left( H_{\rm A}^{\rm qst} ( t ) \right)^2} < 0 .
\end{eqnarray}
From the total differential form of the dimensionless entropy presented in Table~\ref{table:DLPotential}, we obtain $\partial \Theta_{\rm A}^{\rm qst} ( t ) / \partial H_{\rm A}^{\rm qst} ( t ) = \beta^{\rm qst} ( t )$.  Thus, Eq.~\eqref{eq:Concavity} reduces to
\begin{eqnarray}
\frac{\partial \beta^{\rm qst} ( t )}{\partial H_{\rm A}^{\rm qst} ( t )} < 0 ,
\end{eqnarray}
which means that the heat capacity is positive, i.e., $\partial H_{\rm A}^{\rm qst} ( t ) / \partial T^{\rm qst} ( t ) > 0$, with the use of $d \beta^{\rm qst} ( t ) = - {d}T^{\rm qst} ( t ) / k_{\rm B} [T^{\rm qst} ( t ) ]^2$.

For the constant-external-field process, we also have the relation (see Appendix~\ref{EGTS}), expressed as
\begin{eqnarray}
\label{eq:Gibbs-Entropy}
S^{\rm qst}_{\rm A} ( t ) = - \frac{\partial G_{\rm A}^{\rm qst} ( t )}{\partial T^{\rm qst} ( t )},
\end{eqnarray}
where $S_{\rm A}^{\rm qst} ( t ) = k_{\rm B} \Theta_{\rm A}^{\rm qst} ( t )$. Using Eq.~\eqref{eq:Gibbs-Entropy} and the principle of minimum dimensionless work, we can prove that the Gibbs energy is convex for the temperature $T^{\rm qst} ( t )$ (see Appendix~\ref{sec:ConvexGibbs}). Thus, the natural variables of the Gibbs energy are the temperature $T^{\rm qst} ( t )$ and the external field $E^{\rm qst} ( t )$, expressed as $G_{\rm A}^{\rm qst} [ T^{\rm qst} ( t ), E^{\rm qst} ( t ) ]$.

From Eq.~\eqref{eq:LegendreP-E}, we obtain the Legendre transformation for the enthalpy as
\begin{eqnarray}
\label{eq:LegendreH-G}
H_{\rm A}^{\rm qst} ( t ) = G_{\rm A}^{\rm qst} ( t ) + T^{\rm qst} ( t ) S_{\rm A}^{\rm qst} ( t ) .
\end{eqnarray}
Because the Gibbs and Helmholtz energies are related by the Legendre transformation for $E^{\rm qst} ( t )$ and $P_{\rm A}^{\rm qst} ( t )$, the Helmholtz energy is also convex for the temperature $T^{\rm qst} ( t )$.  Thus, from Eq.~\eqref{eq:WA}, we obtain the Legendre transformation between the Helmholtz energy and internal energy as follows:
\begin{eqnarray}
\label{eq:LegendreU-F}
U_{\rm A}^{\rm qst} ( t ) = F_{\rm A}^{\rm qst} ( t ) + T^{\rm qst} ( t ) S_{\rm A}^{\rm qst} ( t ) .
\end{eqnarray}
From Eqs.~\eqref{eq:LegendreG-F},~\eqref{eq:LegendreH-G}, and~\eqref{eq:LegendreU-F}, the thermodynamic potentials are expressed in terms of the natural variables as $F_{\rm A}^{\rm qst} [ T^{\rm qst} ( t ) , P_{\rm A}^{\rm qst} ( t ) ]$, $H_{\rm A}^{\rm qst} [ S_{\rm A}^{\rm qst} ( t ), E^{\rm qst} ( t ) ]$, and $U^{\rm qst}_{\rm A} [ S_{\rm A}^{\rm qst} ( t ), P_{\rm A}^{\rm qst} ( t ) ]$, respectively.

We summarize the thermodynamic potentials as functions of the natural variables in total differential form in Table~\ref{table:Potential}.

Although Table~\ref{table:Potential} is similar to the well-known table of thermodynamics for the equilibrium state, we have obtained it from a dynamical approach using a quasi-static process and have not relied on  calculation of the partition function.  This indicates that the same results can be obtained not only theoretically but also experimentally by measuring heat and polarization in quasi-static processes.

\section{Numerical Demonstrations}
\label{sec:HEOM0}
\subsection{Reduced equations of motion for thermodynamic processes}
\label{sec:HEOM}

We consider the standard Brownian model described as $V(\hat{q})\equiv\hat{q}$ although, if necessary, we can treat nonlinear SB coupling, which is critical for the description of molecular motion.\cite{OT97PRE,TT23JCP1,TT23JCP2} 
We then set $\mu(\hat{q})\equiv \mu \hat{q}$ and introduce the time-dependent potential defined as
\begin{eqnarray}
\label{pote}
U'(\hat q; t) = U(\hat q) - \mu \hat{q} E ( t ) .
\end{eqnarray}

For the reduced density matrix element $\rho_A (q, q';t)=\langle q | \hat \rho_A (t) | q' \rangle$, we introduce the Wigner distribution function, which is the quantum analog of the classical distribution function in phase space, defined as 
\begin{eqnarray}
W_{\rm A} ( p ,q)  \equiv  \frac{1}{{2\pi \hbar }}\int_{ - \infty }^\infty  {dx} \, e^{ipx/\hbar }  \rho_A \left( {q - \frac{x}
{2}, q + \frac{x}
{2}} \right).
\end{eqnarray}
The Wigner distribution function is a real function, in contrast to the complex density matrix: it reduces to the classical distribution function in the classical limit.

We choose the coefficients $\nu _{l}^k$ and $\eta_{l}^k$ in Eq.~\eqref{eq:sym-approx0} for finite $K$. Then, we incorporate this contribution using the HEOM formalism.\cite{T20JCP,IT19JCTC}
%The Matsubara decomposition scheme (MSD) can be applied straightforwardly to the above. In the MSD scheme, we set $\eta _{l}^k=1$ and $\nu _{l}^k=\tilde{\nu }_{l}$, where $\tilde{\nu }_{l}\equiv 2\pi l/\beta \hbar $ is the $l$th Matsubara frequency \cite{T90PRA,IT05JPSJ}.  
To reduce the computational cost, we employ the Pad\'e spectral decomposition (PSD) scheme for $\eta _{l}^k$ and $\nu _{l}^k$ to enhance computational efficiency while maintaining accuracy.\cite{hu2010communication} For time-dependent $\beta(t)$, the decomposition constant becomes a time-dependent function as $\nu _{l}(t)$. 

Under quasi-static conditions, we assume that the time scale of the quantum thermal fluctuations $\beta(t)\hbar/2  \pi$ is shorter than that of the subsystem. Thus, the SB coherence among  different heat baths [e.g., the $k$th bath and the $(k+1)$th bath] is negligible. Under this condition, we can describe the system dynamics using a $K$-dimensional hierarchy instead of a $(K \times N$)-dimensional hierarchy, where $N$ is the number of heat baths.

\subsubsection{Low-temperature quantum Fokker--Planck equations (LT-QFPE) for thermodynamic processes}

Under the PSD scheme, the equations of motion for the Wigner function are expressed as\cite{IT19JCTC}
\begin{eqnarray}
    &&\frac{\partial }{\partial t} {W}_{\vec{n}} (p,q; t) \nonumber \\
    &&=-\left(\mathcal{\hat L}_{qm}(t)+\sum _{l}^{K}n_{l}\nu_{l}(t)+\hat{\Xi }_{K} (p,q; t)\right) {W}_{\vec{n}} (p,q; t) \nonumber \\
    &&\quad -\sum _{l=1}^{K}\hat{\Phi }_p(t) {W}_{\vec{n}+\vec{e}_{l}} (p,q; t) \nonumber \\
    &&\quad -\sum _{l=1}^{K}n_{l}\nu _{l}(t)\hat{\Theta }_{l}(p,q; t) {W}_{\vec{n}-\vec{e}_{l}} (p,q; t),
  \label{eq:lt-qfpe-d}
\end{eqnarray}
where $\vec{n}\equiv (\dots ,n_{k},\dots )$ is a $K$-dimensional multi-index whose components are all non-negative integers and $\vec{e}_{k}\equiv (0,\dots ,1,0,\dots )$ is the $k$th unit vector.
The multi-index $\vec{n}$ represents the index of the hierarchy. Physically, the first hierarchical element, ${W}_{\vec{0}}(p,q,t)$, corresponds to ${W}_{\rm A} (p,q,t)$, and the rest of the hierarchical elements serve only to facilitate the treatment of the non-Markovian system--bath interaction that arises from the hierarchical low-temperature expansion of the noise correlation functions in terms of $e^{- \nu(t)t}$.
In the Wigner representation, the quantum Liouvillian takes the form,\cite{IT19JCTC,Frensley1990}
\begin{eqnarray}
\label{eq:quantumLioiv}
&-& \mathcal{\hat L}_{qm}(t)W (p,\,q) \equiv - \frac{p}{m}\frac{\partial }{{\partial q}}W (p,\,q) \nonumber \\
&+& \sum_{n = 0}^\infty \frac{1}{( 2 n + 1 ) !} \frac{\partial^{2 n + 1} U'(q; t )}{\partial q^{2 n + 1}}
\left( - \frac{\hbar^2}{4} \frac{\partial^2}{\partial p^2} \right)^n \frac{\partial}{\partial p}
W ( p , q ; t ).\nonumber \\
\end{eqnarray}
The operators appearing in Eq.~\eqref{eq:lt-qfpe-d} are defined as
\begin{align}
    \hat{\Phi}_p (t) &\equiv - \frac{A}{\beta ( t )} \frac{\partial }{\partial p},
\end{align}
\begin{align}
\label{Theta}
      \hat{\Theta }_{0}(p,q; t) =\frac{A \beta ( t )}{m} \biggl(p+\frac{m}{\beta(t) }\frac{\partial }{\partial p}\biggr),
\end{align}
\begin{align}
\label{Thetal}
    \hat{\Theta }_{l}(p,q; t)&\equiv 2 A \eta_l    \frac{\partial }{\partial p}~~({\rm for} ~1\le l \le K),
\end{align}
and
\begin{align}
      \hat{\Xi }_{K} (p,q; t)&\equiv \hat{\Phi}_p (t) \sum _{l=0}^{K}\hat{\Theta }_{l}(p,q; t).
\label{Xipq}
\end{align}
Owing to the presence of low-temperature correction terms, the system and bath are entangled,\cite{T20JCP} i.e., $\hat \rho_{\rm tot}(t) \ne \hat \rho_{A}^0(t)\hat \rho_{\rm B}(t)$, where $ \hat \rho_{A}^0 (t) = {\rm tr}_{\rm B} \{ \hat \rho_{\rm tot}(t) \}$. 

Because we consider the case in which the time scale of the quantum thermal fluctuations $\beta(t)\hbar/2  \pi$ is shorter than the time scale of the subsystem $1/\omega_{\rm A}$, the coherence between the subsystem and different heat baths [e.g., the $k$th  and  $(k+1)$th baths] is taken into account by the time-dependent Matsubara frequencies expressed as $\nu_l(t)$.

\subsubsection{Kramers equation for thermodynamic processes}

The Wigner distribution function reduces to the classical distribution function in the limit $\hbar \to 0$, and hence, we can directly compare the quantum results to the classical results.\cite{T15JCP} The classical limit of LT-QFPE is the Kramers equation expressed as\cite{TW91PRA,TW92JCP,IT19JCTC,KRAMERS1940284,risken1996fokker}
\begin{eqnarray}
\frac{\partial}{\partial t}{W} (p, q ; t)&=& -\mathcal{\hat L}_{cl} ( t ) W (p , q ; t) \nonumber \\ 
 &+& \frac{A^2}{m}  \frac{\partial }{\partial p} \left( p + \frac{m}{\beta (t)} \frac{\partial }{\partial p} \right) W(p , q ;t), \nonumber \\
\label{heom_cl}
\end{eqnarray}
where the classical Liouvillian is defined as
\begin{eqnarray}
-\mathcal{\hat L}_{cl}(t)W(p , q) = -\frac{p}{m}\frac{\partial }{\partial q} W(p , q) + \frac{\partial U'(q; t)}{\partial q}\frac{\partial }{\partial p}W(p , q).\nonumber \\
\label{L_cl}
\end{eqnarray}
The description of the Kramers equation is equivalent to that of the Langevin equation\cite{TW91PRA} 
\begin{eqnarray}
m\ddot q + A^2 \dot q + \frac{dU'(q; t)}{dq} +\Omega(t) = 0,
\label{eq:clLangevin}
\end{eqnarray}
with the Gaussian white noise defined as
\begin{eqnarray}
\left\langle {\Omega( t )} \right\rangle = 0,\quad \left\langle {\Omega(t)\Omega(0)} \right\rangle 
=  \frac{A^2 }{\beta ( t )} \delta ( t ).
\label{eq:Langevinforce}
\end{eqnarray}
It should be noted that $W(p, q; t)$ in the Kramers equation is a probability distribution function, whereas $q$ in the Langevin equation is a sampling trajectory and cannot be a distribution function unless the trajectories are averaged over for noise sampling.

Although it is physically impossible to change the temperature of a heat bath with infinite specific heat, Eqs.~\eqref{eq:lt-qfpe-d},~\eqref{heom_cl}, and~\eqref{eq:clLangevin} take  forms in which $\beta$ in the equations of motion derived under the thermostatic process has been replaced with $\beta(t)$. It should be noted, however, that in the non-Ohmic case, the temperature changes even on the time scale on which the noise correlations are defined. This makes it difficult to apply the fluctuation--dissipation theorem to characterize the noise, and thus, a simple replacement $\beta \rightarrow \beta(t)$ is not allowed.

\subsubsection{Isothermal work and constant-external-field heat}

We denote the solution of the reduced density elements obtained from Eq. ~\eqref{eq:lt-qfpe-d} under any $E(t)$ by ${W}_{\vec{n}} (p,q,t)$, whereas that of the classical distribution function obtained from Eq.~\eqref{heom_cl}  is denoted by ${W}(q,p;t)$. The  polarization and enthalpy at time $t$ are evaluated as
\begin{eqnarray}
\label{Mneq3}
P _{\rm A} ( t ) = \mu \left. \mathrm{tr}_{\rm A} \{ q {W}_{\vec{0}} (p,q; t) \} \right|_{T(t)=T} 
\end{eqnarray}
and
\begin{eqnarray}
\label{EnthalpyNumerical}
H_{\rm A} ( t ) = \mathrm{tr}_{\rm A} \left\{ \left[ \frac{p^2}{2 m} + U' ( q ; t ) \right] 
W_{\vec{0}} ( p , q ; t ) \right\} ,
\end{eqnarray}
respectively. By replacing ${W}_{\vec{0}} (p,q,t)$ with ${W} (p,q,t)$ in the above, we can evaluate the classical results. The dimensionless heat current in Eq.~\eqref{DefQ} is expressed as
\begin{eqnarray}
\label{QNumerical1}
\begin{split}
\frac{d Q_{\rm A} ( t )}{d t}
&= - \mathrm{tr}_{\rm A} \left\{ \left( \frac{A p}{m} \right)^2 W_{\vec{0}} ( p , q ; t ) \right\} \\
&
\quad
+ \frac{A^2}{m \beta ( t )} \left( 1 + 2 \sum_{l = 1}^K \eta_l \right) \\
&
\quad
- \sum_{l = 1}^K \mathrm{tr}_{\rm A} \left\{ \frac{A p}{m} W_{\vec{e}_l} ( p , q ; t ) \right\} 
\end{split}
\end{eqnarray}
for the quantum case and as
\begin{eqnarray}
\label{QNumerical2}
\frac{d Q_{\rm A} ( t )}{d t}
= - \mathrm{tr}_{\rm A} \left\{ \left( \frac{A p}{m} \right)^2 W ( p , q ; t ) \right\}
+ \frac{A^2}{m \beta ( t )} 
\end{eqnarray}
for the classical case. However, we find that the accuracy of the numerical result obtained from Eqs.~\eqref{QNumerical1} and~\eqref{QNumerical2} is not sufficient because the heat current $d Q_{\rm A} ( t ) / d t$ does not become zero even when the system approaches equilibrium, and the errors accumulate over a long simulation time. Thus, we calculate the dimensionless heat current using Eqs.~\eqref{Mneq3} and~\eqref{EnthalpyNumerical} as
\begin{eqnarray}
\frac{d \tilde{Q}_{\rm A} ( t )}{d t} = \beta ( t ) \frac{d H_{\rm A} ( t )}{d t}
+ \beta ( t ) P_{\rm A} ( t ) \frac{d E ( t )}{d t} .
\end{eqnarray}
{From Eq.~\eqref{eq:DefQA}, the bath plus SB interaction heat is defined as}
\begin{eqnarray}
\label{DefQIB}
{\frac{d Q_{I + B} ( t )}{d t} = \mathrm{tr}_{\rm tot} \left\{ \hat{H}_{\rm I + B} ( t )
\frac{\partial \hat{\rho}_{\rm tot} ( t )}{\partial t} \right\} }
\end{eqnarray}
{and satisfies $d ( Q_{\rm A} ( t ) + Q_{\rm I + B} ( t ) ) / d t = 0$.}

\subsection{Numerical results}
\label{secSimulations}
\subsubsection{Simulation details}

\begin{figure}[!t]
%\centering%The revtex users' guide advises against the use of explicit centering commands in float environments
\includegraphics[width=6cm]{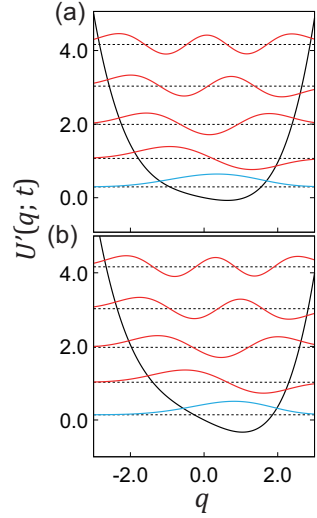}
\caption{\label{FigPotential} Potential surfaces with (a) $E (t = 0) = 0.2$ and (b) $E (t = \tau^{\rm qst}) = 0.5$ are represented by the black curves. 
The blue and red curves represent the ground and excitation states at their respective excitation energies. From this figure, the bath temperature is considered to be low when $T=1/(k_{\rm B}\beta) \approx 0.3$ and high when $T=1/(k_{\rm B}\beta) \approx 5.0$.}
\end{figure}

We perform simulations for an anharmonic potential system to demonstrate that our theory provides a practical means of computing thermodynamic variables and thermodynamic potentials for arbitrary systems.
Thus, we consider a quartic anharmonic potential with  external interaction described by $\mu(\hat q)= \hat q$. The potential function is expressed as
\begin{eqnarray}
\label{SimulationSubsystemHamiltonian}
U'(\hat q) = U_2 \hat{q}^2 + U_3 \hat q^3 + U_4 \hat q^4 - E ( t ) q, 
\end{eqnarray}
where the harmonic and anharmonic constants are $U_2 = 0.1$, $U_3= 0.02$, and $U_4=0.05$. 

Snapshots of the potential surface with the eigenstates and eigenenergies are shown in Fig.~\ref{FigPotential}. 
In the isothermal process, the first excitation energy is $\Delta E_{\rm g \rightarrow e} ( t ) \sim 0.8$, and so the bath temperature is considered low when $T=1/(k_{\rm B}\beta)  \approx 0.3$ and high when $T=1/(k_{\rm B}\beta)  \approx 5.0$.

To obtain a comprehensive theory for thermodynamic potentials, not only the external field $E(t)$ but also the temperature $T(t)$ must be time-dependent as  intensive variables. Thus, we simulate both isothermal and thermostatic processes. 

We consider an isothermal process driven by a quasi-static change in $E(t)$ that consists of  (i) equilibrium, (ii) quasi-static, and (iii) equilibrium steps, defined as
\begin{equation}
\label{IDFprofile}
E^{\rm qst}(t) = \left\{
\begin{array}{rll}
{\rm (i)}  & 0.2 & ( t < 0 ),
\\
{\rm (ii)}  & 0.2 + 0.3 \; t / \tau^{\rm qst} &( 0 \leq t < \tau^{\rm qst} ),
\\
{\rm (iii)} & 0.5 & ( \tau^{\rm qst} \leq t ).
\end{array}
\right.
\end{equation}
Here, the constant $\tau^{\rm qst}$ is the time duration parameter for the quasi-static process, and we set $\tau^{\rm qst} = 1.0 \times 10^4$. 

To {calculate} $S_{\rm A}^{\rm qst} ( t )$, we consider the thermostatic transition {with  fixed external field $E ( t ) = 0.2$} after  isothermal evolution {until time $t = 0$}.
The time profile of $T(t)$ is set by
\begin{equation}
\label{TFprofile}
T^{\rm qst}(t) = \left\{
\begin{array}{rll}
{\rm (i)}  & T_0 &  ( t < 0 ),
\\
{\rm (ii)}  &  ( 1.0 +  t / \tau^{\rm qst} )T_0 &  ( 0 \leq t < \tau^{\rm qst} ),
\\
{\rm (iii)}  & 2 T_0 &  ( \tau^{\rm qst} \leq t ),
\end{array}
\right.
\end{equation}
where $T_0$ is the initial temperature. We perform the simulation for three different cases: $T_0 = 5.0$ (hot), $1.0$ (intermediate), and $0.3$ (cold). 

We summarize the conditions of the numerical simulation for the above-mentioned two cases in Appendix~\ref{sec:NumericalDetails}.

\subsubsection{Results}
\label{sec:NumericalResults}

%\subsubsection{Time evolution in an isothermal process}

\begin{figure}[!t]
\includegraphics[width=5cm]{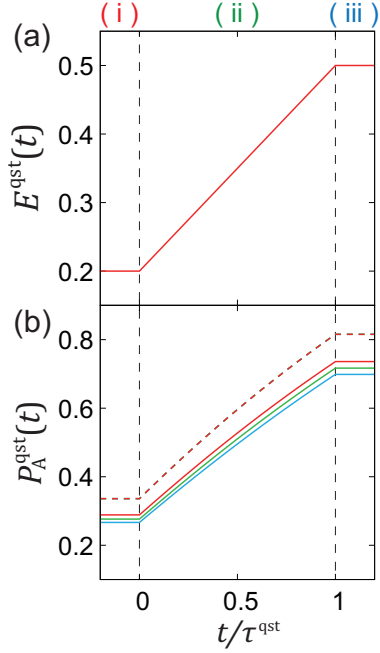}
\caption{\label{EPFig}(a) Time profiles of $E ^{\rm qst}( t )$ for (i) equilibrium, (ii) quasi-static, and (iii) equilibrium steps. (b) Calculated values of $P_{\rm A}(t)$ in the classical case (dashed line) and the quantum case (solid line) for different SB coupling strengths. The red, green, and blue curves represent the strong ($A = 1.5$), intermediate ($A = 1.0$), and weak ($A = 0.5$) SB interaction cases, respectively. The classical results are independent of the SB bond strength, and all overlap.}
\end{figure}

Figure~\ref{EPFig} shows the time profiles of $E^{\rm qst} ( t )$ and the calculated $P_{\rm A}^{\rm qst} ( t )$ in the isothermal case [$T ( t ) = 0.3$]. As $E^{\rm qst} ( t )$ increases, $P_{\rm A}^{\rm qst} ( t )$ also increases because of the conjugate relationship ${\partial G^{\rm qst}_{\rm A} (t)}/{\partial E^{\rm qst}(t)} = P_{\rm A}^{\rm qst}(t)$. 
In the classical case, as can be seen from the steady-state solution of the Kramers equation~\eqref{heom_cl}, we have $Z_{\rm A}^{\rm cl}= \mathrm{tr}_{\rm A} \{ \exp[-\beta(t) \hat{H}_{\rm A} ( t ) ]\}$ and, as in the conventional thermodynamics case, the results are independent of the SB coupling strength. 
In the quantum case, however,  $ P_{\rm A}^{\rm qst}(t)$ becomes smaller with a smaller SB coupling strength. This difference arises from the 
bathentanglement, which is described by the low-temperature correction term in the LT-QFPE. (also see Refs. \onlinecite{Cao2012engangle,Cao2016entangle})
However, as the SB coupling becomes stronger, the system approaches the Smoluchowski limit, where motion is suppressed, and the difference from the classical result becomes smaller.\cite{IT19JCTC}

The change in the Gibbs energy $\Delta G^{\rm qst}_{\rm A}$ in the isothermal process is evaluated from {$( W_{\rm A}^{int} )^{\rm qst}$}. Then, to examine the description of the thermodynamic relation, we compute $\Delta H_{\rm A}^{\rm G}=- T^2 \partial ( \Delta G_{\rm A}^{\rm qst} / T ) / \partial T$ and $\Delta S_{\rm A}^{\rm G}=- \partial \Delta G_{\rm A}^{\rm qst} / \partial T$ and compare these values to the separately calculated values of $\Delta H_{\rm A}^{\rm qst}(t)$ and $\Delta S_{\rm A}^{\rm qst}(t)$ from Eqs.~\eqref{eq:WA} and~\eqref{eq:MaximumDLHeat0}, respectively. The results are presented in Table~\ref{table:DiffGibbs}, from which it can be seen that Eqs.~\eqref{eq:GibbsHelmholtz2} and~\eqref{eq:GSRelation2} are valid

\begin{table}[!t]
\caption{\label{table:DiffGibbs} Enthalpy and entropy changes and  temperature derivative of the Gibbs energy for each interaction strength $A$. Here, $\Delta H_{\rm A}^{\rm G}$ and $\Delta S_{\rm A}^{\rm G}$ are calculated from Eqs.~\eqref{eq:GibbsHelmholtz2} and~\eqref{eq:GSRelation2}, respectively. The results in this table indicate that the relation $\Delta G_{\rm A}^{\rm qst} = \Delta H_{\rm A}^{\rm G} + T^{\rm qst} \Delta S_{\rm A}^{\rm G}$ is satisfied.}
\begin{ruledtabular}
\begin{tabular}{cccccc}
$A$ & $\Delta H_{\rm A}^{\rm qst}$ & $\Delta H_{\rm A}^{\rm G}$
& $\Delta S_{\rm A}^{\rm qst}$ & $\Delta S_{\rm A}^{\rm G}$ & $\Delta G_{\rm A}^{\rm qst}$
\\
\hline
$0.5$ & $-0.170$ & $-0.170$ & $-7.63 \times 10^{-2}$ & $-7.62 \times 10^{-2}$ & $-0.147$
\\
$1.0$ & $-0.181$ & $-0.180$ & $-9.81 \times 10^{-2}$ & $-9.50 \times 10^{-2}$ & $-0.152$
\\
$1.5$ & $-0.192$ & $-0.192$ & $-0.117$ & $-0.116$ & $-0.157$
\end{tabular}
\end{ruledtabular}
\end{table}

\begin{figure}[!t]
\includegraphics[width=5cm]{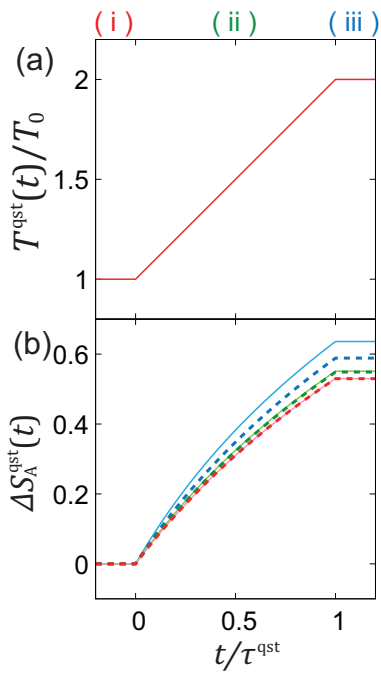}
\caption{\label{TSFig} (a) Time profiles of $\beta^{\rm qst}( t )$ for (i) equilibrium, (ii) quasi-static, and (iii) equilibrium steps. (b)
Time profile of $S_{\rm A}^{\rm qst}(t)$ for different temperatures {in the classical case (dashed line) and the quantum case (solid line) for different SB coupling strengths.} The red, green, and blue curves represent the cases of high ($T_0 = 5.0$), intermediate ($T_0 = 1.0$), and low ($T_0 = 0.3$) bath temperatures, respectively. {The classical and quantum results approach each other at higher temperatures and  almost overlap for the intermediate- and high-temperature cases.}
}
\end{figure}

Figure~\ref{TSFig} presents the time profiles of $T ( t )$ and the change in entropy $\Delta S_{\rm A}^{\rm qst}(t)$ as a function of $t$ in the constant-external-field case $A = 1.0$. Because $S_{\rm A}^{\rm qst}(t)$ satisfies the conjugate relation ${\partial G^{\rm qst}_{\rm A} (t)}/{\partial T^{\rm qst}(t)} = - S_{\rm A}^{\rm qst}(t)$, the variation of $S_{\rm A}^{\rm qst}(t)$ in time is similar to that of $P_{\rm A}^{\rm qst}(t)$, while $S_{\rm A}^{\rm qst}(t)$ does not increase linearly with $t$ because the role of $T(t)$ in ${G^{\rm qst}_{\rm A} ( t )}$ is completely different from $E(t)$. At low temperatures, quantum bathentanglement becomes important, and the differences between classical and quantum results are pronounced. At high and intermediate temperatures $T_0 = 5.0$ and $1.0$, on the other hand, the contribution from the Matsubara frequency terms becomes smaller, and so the quantum result almost overlaps with the classical result.

\section{Conclusions}
\label{sec:conclude}
The virtue of thermodynamics lies in its ability to describe macroscopic thermal phenomena resulting from complex microscopic interactions in a system-independent manner as changes in thermodynamic potentials, which are described as interrelated intensive and extensive variables using Legendre transformations. This virtue should be preserved when we develop a quantum thermodynamic theory rather than an open quantum dynamical theory, although in either case, the theory must be specific to the SB model.
Moreover, as a first-principles argument based on the SB model, the thermodynamic laws themselves should be derived within the framework of open quantum dynamical theory. Thus, in the present study, we have demonstrated the following:
\begin{enumerate}
\item A model consisting of multiple heat baths at different temperatures (thermodynamic SB model) has been introduced to describe the thermostatic process. Work and heat have then been defined as the expectation values of the Hamiltonian system. 
\item Extensive variables, such as polarization and enthalpy, {have been defined as physical variables that can describe not only equilibrium but also non-equilibrium processes.} 
\item Dimensionless (or entropic) work and heat satisfy the time-dependent (non-equilibrium) Legendre transformations as extensive variables and their conjugate intensive variables.
\item We evaluate the Massieu and Planck potentials using the minimum value of the dimensionless extensive and intensive work, respectively, and we evaluate the Boltzmann entropy using the maximum value of the dimensionless heat (the principles of dimensionless minimum work and dimensionless maximum entropy).  Expressions for the entropic potentials in  total differential form as functions of natural variables are presented in Table~\ref{table:DLPotential}. 
\item These principles are a consequence of the fact that the number of energy states (thermodynamic weights) is maximized (or entropy is maximized) when the distribution of states becomes a canonical ensemble for the energy given at that instant, even if the energy of the subsystem changes with time due to temperature changes in the heat bath.
\item The Gibbs and Helmholtz energies, enthalpy, and internal energy can be evaluated from entropic potentials. The expressions for these functions in  total differential form as functions of natural variables are presented in Table~\ref{table:Potential}.
\item The first and second laws of thermodynamics also follow naturally from these arguments.
\item Numerically accurate simulations in both classical and quantum cases have been performed using the reduced equations of motion developed for the thermodynamic SB model. 
\item The differences between the classical and quantum results can be attributed to quantum entanglement between the system and bath (i.e., bathentanglement).
\end{enumerate}

The above-mentioned statements indicate that the model and formalism introduced in the present study are powerful means for accurate simulation and analysis of thermodynamic problems in both classical and quantum regimes. If the limitations due to computational cost can be removed,  it will be possible to investigate the thermodynamic properties of any system as a subsystem, such as molecular liquids and proteins using the Langevin formalism as classical examples,\cite{IIT15JCP} and spin-lattice\cite{NT18PRA} and superconductor systems.\cite{NT21JCP}

Finally, using the present model, the definitions of non-equilibrium intensive and extensive variables, and the reduced equations of motion, it is possible to study thermodynamic processes in a fully non-equilibrium regime. Thus, we will show in a subsequent paper that the thermodynamic relationships presented in Tables~\ref{table:DLPotential} and~\ref{table:Potential} can be systematically extended to the non-equilibrium regime in terms of the time-dependent intensive variables $E(t)$ and $T(t)$ and extensive variables $P_{\rm A}(t)$ and $S_{\rm A}(t)$ by introducing waste heat.\cite{KT24JCP2}

\section*{Acknowledgments}
Y. T. thanks Peter Wolynes for pointing out the relationship between the dimensionless thermodynamic potentials and the Massiue--Planck potentials.
Y.T. was supported by JSPS KAKENHI (Grant No.~B21H01884).
S.K. acknowledges a fellowship supported by JST, the establishment of university fellowships toward the creation of science technology innovation (Grant No.~JPMJFS2123). S. K. was also supported by Grant-in-Aid for JSPS Fellows (Grant No.~24KJ1373).

\section*{Author declarations}
\subsection*{Conflict of Interest}
The authors have no conflicts to disclose.

%\subsection*{Author Contributions}

\section*{Data availability}
The data that support the findings of this study are available from the corresponding author upon reasonable request.

\appendix

\section{Dimensionless minimum work principle}
\label{sec:MinimumWorkProof}

To obtain the dimensionless minimum work principle for multiple-bath systems, the derivation of the Jarzynski equality  is modified.\cite{tasaki2000, ST20JCP}

We assume that the subsystem is initially connected only to the first bath ($k=1$), and the total density operator at time $t_0$ is expressed as
\begin{eqnarray}
\label{eq:Jarzynski1}
\hat{\rho}_{\rm tot}^{\rm init}
= \frac{1}{Z_{\rm tot}^{\rm init}}
 e^{- \hat{\tilde{H}}_{\rm tot}(t_0)},
\end{eqnarray}
where $\hat{\tilde{H}}_{\rm tot} ( t ) $ is defined in Eq.~\eqref{eq:DefTHtot} and
$Z_{\rm tot}^{\rm init}$ is the total partition function expressed as
\begin{eqnarray}
\label{eq:Jarzynski2}
Z_{\rm tot}^{\rm init} ( t_0 ) =
Z_{\rm A + IB}^1 ( t_0 ) \prod_{k = 2}^N Z_{\rm B}^k, 
\end{eqnarray}
with
\begin{eqnarray}
\label{eq:Jarzynski2-1}
Z_{\rm A + IB}^k ( t ) = \mathrm{tr}_{\mathrm{A + I}\mathrm{B}^k} 
\left\{  e^{ - \beta_k \left( \hat{H}_{\rm A} ( t ) + \hat{H}_{\rm I + B}^k ( t ) \right)} \right\} 
\end{eqnarray}
and
\begin{eqnarray}
\label{eq:Jarzynski2-2}
Z_{\rm B}^k = \mathrm{tr}_{\mathrm{B}^k} \left\{ e^{- \beta_k \hat{H}_{\rm B}^k} \right\} .
\end{eqnarray}
Under the assumption that only the $N$th bath as being connected to the subsystem at a time $t$, the partition function of the final state is given by 
\begin{eqnarray}
\label{eq:Jarzynski4}
Z^{\rm fin}_{\rm tot} = Z_{\rm A + IB}^N ( t ) \prod_{k = 1}^{N - 1} Z_{\rm B}^k.
\end{eqnarray}
The time evolution of the total system from $t_0$ to $t$ is described by the operator $\hat{\mathcal{G}} ( t , t_0 )$. We then have 
\begin{multline}
\label{eq:Jarzynski3}
\frac{Z_{\rm A} ( t )}{Z_{\rm A} ( t_0 )}  
= \\
\mathrm{tr}_{\rm tot}\left\{ \hat{\mathcal{G}}^\dagger_{\rm tot} ( t , t_0 )
 e^{- \hat{\tilde{H}}_{\rm tot}(t)}\hat{\mathcal{G}}_{\rm tot} ( t , t_0 )    e^{\hat{\tilde{H}}_{\rm tot} ( t_0 )}
\hat{\rho}_{\rm tot}^{\rm init} \right\}, 
\end{multline}
where $Z_{\rm A} ( t ) = Z_{\rm A + IB}^k / Z_{\rm B}^k$ [for $\xi_k ( t ) \neq 0$] is the partition function of the subsystem.

Let $| \tilde{D}_i \rangle$ and $| \tilde{E}_j \rangle$ be the eigenkets of the operators $\hat{\tilde{H}}_{\rm tot}(t_0)$ and $\hat{\tilde{H}}_{\rm tot}(t)$.  The right-hand side of Eq.~\eqref{eq:Jarzynski3} is then expressed as
\begin{eqnarray}
\label{eq:Jarzynski6}
\sum_{i , j} e^{- ( \tilde{E}_j - \tilde{D}_i )} | \langle \tilde{E}_j | \hat{\mathcal{G}}_{\rm tot} ( t , t_0 ) | \tilde{D}_i \rangle |^2
p ( \tilde{D}_i ) ,
\end{eqnarray}
where {$p ( \tilde{D}_i ) = \langle \tilde{D}_i | \exp ( -\hat{\tilde{H}}_{\rm tot} (t_0) ) | \tilde{D}_i \rangle / Z_{\rm tot}^{\rm init}$} is the population of the $i$th state. 
Applying Jensen's inequality to Eq.~\eqref{eq:Jarzynski6}, we obtain
\begin{eqnarray}
\label{eq:Jarzynski7}
\frac{ e^{- \langle \hat{\tilde{H}}_{\rm tot}(t)
\rangle }}
{ e^{- \langle \hat{\tilde{H}}_{\rm tot}(t_0) \rangle  }
}  \leq \frac{Z_{\rm A} ( t )}{Z_{\rm A} ( t_0 )}.
\end{eqnarray}
Taking the logarithm of both sides of Eq.~\eqref{eq:Jarzynski7}, we have the inequality
\begin{eqnarray}
\label{eq:Jarzynski8}
\int^t_{t_0} \frac{\partial}{\partial t'} \mathrm{tr}_{\rm tot}
\left\{ \hat{\tilde{H}}_{\rm tot}(t')
\hat{\rho}_{\rm tot} ( t' ) \right\} d t' \geq - \Delta \Xi_{\rm A} ,
\end{eqnarray}
where $\Xi_{\rm A} ( t ) = \ln Z_{\rm A} ( t )$ is the Planck potential. 

When the $k$th bath is connected to and disconnected from the subsystem, we have 
\begin{eqnarray}
\label{eq:Jarzynski8-1}
\mathrm{tr}_{\rm tot} \left\{ \left[ \hat{H}_{\rm A} ( t ) + \hat{H}_{\rm I + B} ( t ) \right]
\frac{\partial \hat{\rho}_{\rm tot} ( t )}{\partial t} \right\} = 0 
\end{eqnarray}
and
\begin{eqnarray}
\label{eq:Jarzynski8-2}
\mathrm{tr}_{\rm tot} \left\{ \hat{H}_{\rm B}^k
\frac{\partial \hat{\rho}_{\rm tot} ( t )}{\partial t} \right\} = 0 .
\end{eqnarray}
Thus, the left-hand side of Eq.~\eqref{eq:Jarzynski8} reduces to
\begin{eqnarray}
\label{eq:Jarzynski9}
\int^t_{t_0} 
\mathrm{tr}_{\rm tot} \left\{ \frac{\partial \hat{\tilde{H}}_{\rm tot}(t')}{\partial t'}  \hat{\rho}_{\rm tot} ( t' ) \right\}
d t' .
\end{eqnarray}
{In the case where} the work due to the SB interaction is incorporated into the system.
Thus, we obtain the dimensionless {(or entropic)} minimum work principle expressed as
\begin{eqnarray}
\label{eq:Jarzynski10}
\int^t_{t_0} \mathrm{tr}_{\rm A} \left\{ \frac{\partial}{\partial t'} \left[ \beta ( t' ) \hat{H}_{\rm A} ( t' )  \right]
\hat{\rho}_{\rm A} ( t' ) \right\} \geq - \Delta \Xi_{\rm A} .
\end{eqnarray}

\section{Equality of the minimum work principle}
\label{sec:qstProof}

In this appendix, we consider only the quasi-static case and omit the superscript $k$ in Eqs.~\eqref{eq:SBHamiltonian}--\eqref{BA}. When $J ( \omega )$ is Ohmic, the spectral density is invariant under the transformation $\hat{H}_{\rm I + B} \rightarrow a \hat{H}_{\rm I + B}$ with $m_j \rightarrow m_j / a$ , $\omega_j \rightarrow a \omega_j$, and $c_j \rightarrow a c_j$,
 where $a > 0$ is a dimensionless scaling parameter.

Because the dynamics of the subsystem depend only on temperature, $J ( \omega )$,  $\hat{H}_{\rm A} ( t )$, and 
$\hat{\rho}_{\rm A} ( t )$ in the quasi-static process are independent of the choice of $a$. Therefore, we can set the parameter $a = \beta_0 / \beta ( t )$ for later convenience without loss of generality, where $\beta_0 > 0$ is the arbitrary inverse temperature.

Using Kubo's identity, we have\cite{ST20JCP}
\begin{equation}
\label{KuboIdentity}
\begin{split}
&  e^{- \beta ( t + \Delta t ) \hat{H}_{\rm tot} ( t + \Delta t )} 
- e^{- \beta ( t ) \hat{H}_{\rm tot} ( t )}
\\
&
= - \int^1_0 d \lambda e^{- ( 1 - \lambda ) \beta ( t ) \hat{H}_{\rm tot} ( t )}
\\
& \quad \times
\left\{ \beta ( t + \Delta t ) \hat{H}_{\rm tot} ( t + \Delta t )
- \beta ( t ) \hat{H}_{\rm tot} ( t ) \right\}
\\
& \quad \times
e^{- \lambda \beta ( t + \Delta t ) \hat{H}_{\rm tot} ( t + \Delta t )} .
\end{split}
\end{equation}
Expanding Eq.~\eqref{KuboIdentity} in $\Delta t$ and taking the limit $\Delta t \rightarrow 0$, we obtain 
\begin{equation}
\label{KuboIdentity2}
\begin{split}
\frac{\partial}{\partial t} e^{- \beta ( t ) \hat{H}_{\rm tot} ( t )}
 ={}& - \int^1_0 d \lambda e^{- ( 1 - \lambda ) \beta ( t ) \hat{H}_{\rm tot} ( t )} 
\\
& \times
\frac{\partial}{\partial t} \left[ \beta ( t ) \hat{H}_{\rm A} ( t ) \right]
e^{- \lambda \beta ( t ) \hat{H}_{\rm tot} ( t )} .
\end{split}
\end{equation}
Here, we have employed the equality,
\begin{eqnarray}
\frac{\partial}{\partial t} \left[ \beta ( t ) \hat{H}_{\rm tot} ( t )  \right]
= \frac{\partial}{\partial t} \left[ \beta ( t ) \hat{H}_{\rm A} ( t )  \right] ,
\end{eqnarray}
using the fact that $\beta ( t ) \hat{H}_{\rm I + B} ( t )$ is time-independent because the time-dependent terms in $\beta ( t )$ and $\hat{H}_{\rm I + B} ( t )$ cancel out. Taking the trace of the total system on both sides of Eq.~\eqref{KuboIdentity2}, we obtain 
\begin{equation}
\label{KuboIdentity3}
\frac{\partial}{\partial t} Z_{\rm A} ( t )
= - \mathrm{tr}_{\rm tot} \left\{ \frac{\partial}{\partial t} \left[ \beta ( t ) \hat{H}_{\rm A} ( t ) \right]
\frac{e^{- \beta ( t ) \hat{H}_{\rm tot} ( t )}}{Z_{\rm B}} \right\} ,
\end{equation}
where  $Z_{\rm A} ( t ) = \mathrm{tr}_{\rm tot} \{ e^{- \beta ( t ) \hat{H}_{\rm tot} ( t )} \} / Z_{\rm B}$ is the partition function of the subsystem and $Z_{\rm B} = \mathrm{tr}_{\rm B} \{ e^{- \beta ( t ) \hat{H}_{\rm B} ( t )} \}$ is the bath partition function, which is time-independent. Dividing both sides of Eq.~\eqref{KuboIdentity3} by $Z_{\rm A} ( t )$, we obtain the equality,
\begin{equation}
\label{qstEquality}
\frac{\partial}{\partial t} \left[ \beta ( t ) G_{\rm A} ( t ) \right]
= \mathrm{tr}_{\rm A} \left\{ \frac{\partial}{\partial t} \left[ \beta ( t ) \hat{H}_{\rm A} ( t ) \right]
\hat{\rho}_{\rm A}^{\rm qst} ( t ) \right\} ,
\end{equation}
where we have introduced the quasi-static Gibbs energy $G_{\rm A} ( t ) = - \ln Z_{\rm A} ( t ) / \beta ( t )$ and the reduced density operator in the quasi-static process $\hat{\rho}_{\rm A}^{\rm qst} ( t )$. 
Solving Eq.~\eqref{qstEquality} for $d G_{\rm A} ( t ) / d t$, we obtain 
\begin{align}
\label{Gdiff}
\frac{d G_{\rm A}^{\rm qst} ( t )}{d t}
={}& - \frac{\mathrm{tr}_{\rm A} \{ \hat{H}_{\rm A} ( t ) \hat{\rho}_{\rm A}^{\rm qst} ( t ) \} - G_{\rm A}^{\rm qst} ( t )}
{T^{\rm qst} ( t )} 
\frac{d T^{\rm qst} ( t )}{d t} \nonumber\\
& 
+ \mathrm{tr}_{\rm A} \{ \mu ( \hat{q} ) \hat{\rho}_{\rm A}^{\rm qst} ( t ) \} \frac{d E^{\rm qst} ( t )}{d t} .
\end{align}
By comparing Eq.~\eqref{Gdiff} with $d G_{\rm A}^{\rm qst} = - S_{\rm A}^{\rm qst} d T^{\rm qst} + P^{\rm qst}_{\rm A} d E^{\rm qst}$, the entropy and polarization are evaluated in terms of the partition function as
\begin{equation}
S_{\rm A}^{\rm qst} ( t ) = k_{\rm B} \left[ \beta^{\rm qst} ( t ) 
\mathrm{tr}_{\rm A} \{ \hat{H}_{\rm A} ( t ) \hat{\rho}_{\rm A}^{\rm qst} ( t ) \}
- \ln Z_{\rm A} ( t ) \right]
\end{equation}
and
\begin{eqnarray}
P^{\rm qst}_{\rm A} ( t ) = \mathrm{tr}_{\rm A} \{ \mu ( \hat{q} ) \hat{\rho}_{\rm A}^{\rm qst} ( t ) \} .
\end{eqnarray}

\section{Convexity and concavity of thermodynamic functions}
\label{sec:Convexity}

\begin{figure}[!t]
\includegraphics[width=7cm]{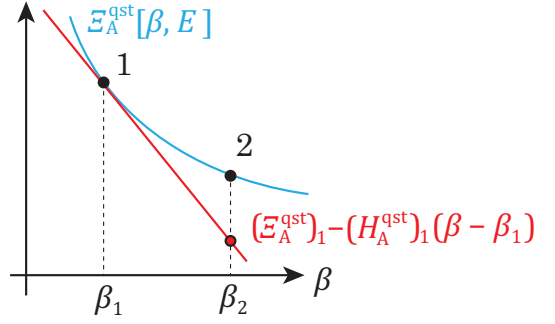}
\caption{\label{FigConvex} Schematic  illustration of the convexity of $\Xi_{\rm A}^{\rm qst}$. The blue curve represents the Planck potential as a function of the inverse temperature $\beta$ for fixed $E=E_{\rm fix}$.  The red line represents the tangent line of the blue curve at the inverse temperature $\beta_1$, expressed as $(\Xi_{\rm A}^{\rm qst})_1 - (H_{\rm A}^{\rm qst})_1 ( \beta - \beta_1 )$. The red line is below the blue curve.} 
\end{figure}

We show the convexity of the Planck potential for $\beta^{\rm qst} ( t )$.  For a fixed external field ${E ( t )=E_{\rm fix}}$, consider the two equilibrium states $\alpha = 1$ at time $t_1$ and $\alpha = 2$ at time $t_2$. 
Under this condition, the dimensionless minimum work principle is expressed as
\begin{eqnarray}
\label{eq:Convex1}
- \int^{t_2}_{t_1} H_{\rm A} ( t' ) \frac{d \beta ( t' )}{d t'} d t' 
\leq \left(\Xi_{\rm A}^{\rm qst} \right)_2 - \left(\Xi_{\rm A}^{\rm qst}\right)_1 ,
\end{eqnarray}
where $\left(\Xi_{\rm A}^{\rm qst} \right)_{\alpha}$ is the Planck potential for $\alpha = 1$ or 2.
We then apply the inequality~\eqref{eq:Convex1} for the thermostatic process as follows:
\begin{equation}
\label{eq:Convex2}
\beta ( t ) = 
\begin{cases}
\beta_1 & ( t \leq t_1 ),
\\
\beta_2 & ( t > t_1 ).
\end{cases}
\end{equation}
The left-hand side of Eq.~\eqref{eq:Convex1} can be evaluated as $- H_{\rm A} ( t_1 ) ( \beta_2 - \beta_1 )$. 
Because the system is in equilibrium at $t_1$, the enthalpy is expressed in terms of its equilibrium value as $H_{\rm A} ( t_1 ) = (H_{\rm A}^{\rm qst})_1$. Using $H_{\rm A}^{\rm qst} ( t ) = - \partial \Xi_{\rm A}^{\rm qst} ( t ) / \partial \beta^{\rm qst} ( t )$ for  Eq.~\eqref{eq:Convex1}, we obtain 
\begin{eqnarray}
\label{eq:Convex3}
\left(\Xi_{\rm A}^{\rm qst} \right)_1 + \left.  \frac{\partial \Xi_{\rm A}^{\rm qst} [ \beta , E_{\rm fix} ]}{\partial \beta} 
\right|_{\beta = \beta_1} ( \beta_2 - \beta_1 ) \leq  \left(\Xi_{\rm A}^{\rm qst} \right)_2 .
\end{eqnarray}
As shown in Fig.~\ref{FigConvex}, the left-hand side of Eq.~\eqref{eq:Convex3} corresponds to the tangent line of the Planck potential. The inverse temperature $\beta_1$ and $\beta_2$ are arbitrary, indicating that the Planck potential is convex with respect to $\beta^{\rm qst} ( t )$.

In the same manner and from the property of the Legendre transformation, we can show the convexity and concavity of the dimensionless entropic potentials. We summarize the results in Table~\ref{table:Convex}.

\begin{table}[!t]
\caption{\label{table:Convex} Convexity and concavity of the entropic potentials for the natural variables.}
\begin{ruledtabular}
\begin{tabular}{lcll}
Potential & Symbol & \multicolumn{2}{c}{Convexity and concavity} 
\\
\hline 
Massieu & $\Phi_{\rm A}^{\rm qst}$ & $\beta^{\rm qst}$: convex & $\tilde{P}_{\rm A}^{\rm qst}$: concave
\\
Planck & $\Xi_{\rm A}^{\rm qst}$ & $\beta^{\rm qst}$: convex & $E^{\rm qst}$: convex
\\
B-entropy & $\Theta_{\rm A}^{\rm qst}$ & $H_{\rm A}^{\rm qst}$: concave & $E^{\rm qst}$: convex
\\
C-entropy & $\Pi_{\rm A}^{\rm qst}$ 
& $H_{\rm A}^{\rm qst}$: concave & $\tilde{P}_{\rm A}^{\rm qst}$: concave
\end{tabular}
\end{ruledtabular}
\end{table}

\section{Internal energy representation}
\label{sec:InternalRep}

For the description of the entropic potentials, here we employ the internal energy instead of the enthalpy. From Table~\ref{table:DLPotential}, we have
\begin{eqnarray}
\label{eq:DiffP-H}
d \Xi_{\rm A}^{\rm qst}  = - H_{\rm A}^{\rm qst} d \beta^{\rm qst} + \tilde{P}_{\rm A}^{\rm qst} d E^{\rm qst}.
\end{eqnarray}
Substituting Eq.~\eqref{eq:LegendreH-U} into this, the Planck potential is expressed as
\begin{eqnarray}
\label{eq:DiffP-U}
d \Xi_{\rm A}^{\rm qst} ( t ) = - U_{\rm A}^{\rm qst} ( t ) d \beta^{\rm qst} ( t )
+ P_{\rm A}^{\rm qst} ( t ) d \tilde{E}^{\rm qst} ( t ) ,
\end{eqnarray}
where $\tilde{E}^{\rm qst} ( t ) = \beta^{\rm qst} ( t ) E^{\rm qst} ( t )$. Applying the Legendre transformation to this, we obtain the Massieu potential and B-entropy in the internal energy representation as 
\begin{eqnarray}
\label{eq:DiffM-U}
d \Phi_{\rm A}^{\rm qst} ( t ) = - U_{\rm A}^{\rm qst} ( t ) d \beta^{\rm qst} ( t )
- \tilde{E}^{\rm qst} ( t ) d P_{\rm A}^{\rm qst} ( t )
\end{eqnarray}
and
\begin{eqnarray}
\label{eq:DiffL-U}
d \Theta_{\rm A}^{\rm qst} ( t ) = \beta^{\rm qst} ( t ) d U_{\rm A}^{\rm qst} ( t )
- \tilde{E}^{\rm qst} ( t ) d P_{\rm A}^{\rm qst} ( t ) ,
\end{eqnarray}
where we have used $E^{\rm qst} ( t ) \tilde{P}_{\rm A}^{\rm qst} ( t ) = \tilde{E}^{\rm qst} ( t ) P_{\rm A}^{\rm qst} ( t )$ and $\beta^{\rm qst} ( t ) H_{\rm A}^{\rm qst} ( t ) = \beta^{\rm qst} ( t ) U_{\rm A}^{\rm qst} ( t ) - \tilde{E}^{\rm qst} ( t )  P_{\rm A}^{\rm qst} ( t )$. In Eq.~\eqref{eq:DiffL-U}, the B-entropy is expressed in terms of extensive variables, while the value of B-entropy is unchanged. However, in the internal energy representation, we cannot obtain the C-entropy from the Legendre transformation of the B-entropy, because the sign of the second term in Eq.~\eqref{eq:DiffL-U} is opposite to that presented in Table~\ref{table:DLPotential}.

\section{Enthalpy and Gibbs energy as functions of $T^{\rm qst}(t)$ and $S^{\rm qst}(t)$}
\label{EGTS}

Here, we consider the fixed-external-field process [$E ( t ) = E_{\rm fix}$].  From the total differential form of the Planck potential presented in Table~\ref{table:DLPotential},  we have
\begin{eqnarray}
\label{eq:GibbsHelmholtz0}
\frac{d \Xi_{\rm A}^{\rm qst} ( t )}{d t} = - H_{\rm A}^{\rm qst} ( t ) \frac{d \beta^{\rm qst} ( t )}{d t} .
\end{eqnarray}
Because the Gibbs energy is defined as $G_{\rm A}^{\rm qst} ( t ) = - \Xi_{\rm A}^{\rm qst} ( t ) / \beta$ for  arbitrary  $\beta$, we can replace $\beta$ with $\beta^{\rm qst} ( t )$.  In this way, we can extend this definition of the Gibbs energy to the thermostatic process.  Then, substituting $\Xi_{\rm A}^{\rm qst} ( t ) = {- \beta^{\rm qst} ( t )} G_{\rm A}^{\rm qst} ( t )$ in Eq.~\eqref{eq:GibbsHelmholtz0}, we have
\begin{eqnarray}
\label{eq:GibbsHelmholtz1}
\frac{d}{d t} \left[ \beta^{\rm qst} ( t ) G_{\rm A}^{\rm qst} ( t ) \right] = H_{\rm A}^{\rm qst} ( t ) 
\frac{d \beta^{\rm qst} ( t )}{d t},
\end{eqnarray}
which leads to
\begin{eqnarray}
\label{eq:GibbsHelmholtz2}
H_{\rm A}^{\rm qst} ( t ) = \frac{\partial}{\partial \beta^{\rm qst} ( t )} 
\left[ \beta^{\rm qst} ( t ) G_{\rm A}^{\rm qst} ( t ) \right] .
\end{eqnarray}
Equation~(\ref{eq:GibbsHelmholtz1}) can then be rewritten as
\begin{eqnarray}
\label{eq:GSRelation1}
\frac{d G_{\rm A} ( t )}{d t} = 
\frac{ \Xi_{\rm A}^{\rm qst} ( t ) + \beta^{\rm qst} ( t ) H_{\rm A}^{\rm qst} ( t )}{ ( \beta^{\rm qst} ( t ) )^2}
\frac{d \beta^{\rm qst} ( t )}{d t} ,
\end{eqnarray}
and thus we have
\begin{eqnarray}
\label{eq:GSRelation2}
S_{\rm A} ( t ) =
k_{\rm B} ( \beta^{\rm qst} ( t ) )^2 \frac{\partial G_{\rm A} ( t )}{\partial \beta^{\rm qst} ( t )},
\end{eqnarray}
where we have applied the Legendre transformation in Eq.~\eqref{eq:LegendreP-E} to the right-hand side of Eq.~\eqref{eq:GSRelation1}.

\section{Concavity of the Gibbs energy}
\label{sec:ConvexGibbs}

Here, we prove the concavity of the Gibbs energy for $T^{\rm qst} ( t )$ for a fixed external field $E^{\rm qst} ( t )=E_{\rm fix}$.
We then consider the transition from equilibrium state 1 to another equilibrium state 2, as given in Appendix~\ref{sec:Convexity}, where the inequality~\eqref{eq:Convex1} holds. For the thermostatic process defined in Eq.~\eqref{eq:Convex2}, we obtain the inequality expressed as
\begin{eqnarray}
\label{eq:GConvex1}
-\left(H_{\rm A}^{\rm qst}\right)_1 ( \beta_2 - \beta_1 ) \leq 
\left(\Xi_{\rm A}^{\rm qst} \right)_2- \left(\Xi_{\rm A}^{\rm qst} \right)_1 .
\end{eqnarray}
Then, using the equality $(H_{\rm A}^{\rm qst})_1 = ( (\Theta_{\rm A}^{\rm qst})_1 - (\Xi_{\rm A}^{\rm qst} )_1 ) / \beta_1$, which we obtain from the Legendre transformation for the dimensionless B-entropy in the quasi-static (equilibrium) state 1 (i.e., $(\Theta_{\rm A}^{\rm qst})_1 = \beta_1 ( H_{\rm A }^{\rm qst} )_1 + (\Xi_{\rm A}^{\rm qst} )_1$ ), we obtain 
\begin{eqnarray}
\label{eq:GConvex2}
 \left( \Theta_{\rm A}^{\rm qst} \right)_1 \left( \frac{\beta_2}{\beta_1} - 1 \right)
\leq \left(\Xi_{\rm A}^{\rm qst} \right)_2 - \frac{\beta_2}{\beta_1} \left(\Xi_{\rm A}^{\rm qst} \right)_1 .
\end{eqnarray}
Dividing both sides of Eq.~\eqref{eq:GConvex2} by $- \beta_2$, we have the inequality for the Gibbs energy as
\begin{eqnarray}
\label{eq:GConvex3}
- \left(S_{\rm A}^{\rm qst} \right)_1 ( T_2 - T_1 ) \geq \left(G_{\rm A}^{\rm qst}\right)_2 - \left( G_{\rm A}^{\rm qst}\right)_1,
\end{eqnarray}
where we have used the definitions of the quasi-static entropy $(S_{\rm A}^{\rm qst})_1  = k_{\rm B} (\Theta_{\rm A}^{\rm qst})_1$ and of the quasi-static Gibbs energy $(G_{\rm A}^{\rm qst})_{\alpha} = -(\Xi_{\rm A}^{\rm qst})_{\alpha} / \beta_\alpha \; ( \alpha = 1 , 2 )$. Because we have Eq.~\eqref{eq:Gibbs-Entropy}, Eq.~\eqref{eq:GConvex3} leads to the concavity of the Gibbs energy for the temperature.

We summarize the convexity and concavity of the thermodynamic potentials in Table~\ref{table:ConvexG}.

\begin{table}[!t]
\caption{\label{table:ConvexG} Convexity and concavity of the thermodynamic potentials for the natural variables.}
\begin{ruledtabular}
\begin{tabular}{lcll}
Potential & Symbol & \multicolumn{2}{c}{Convexity and concavity} 
\\
\hline
Helmholtz & $F_{\rm A}^{\rm qst}$ & $T^{\rm qst}$: concave & $\tilde{P}_{\rm A}^{\rm qst}$: convex
\\
Gibbs & $G_{\rm A}^{\rm qst}$ & $T^{\rm qst}$: concave & $E^{\rm qst}$: concave
\\
Internal energy & $U_{\rm A}^{\rm qst}$ & $S_{\rm A}^{\rm qst}$: convex & $\tilde{P}_{\rm A}^{\rm qst}$: convex
\\
Enthalpy & $H_{\rm A}^{\rm qst}$ & $S_{\rm A}^{\rm qst}$: convex & $E^{\rm qst}$: concave
\end{tabular}
\end{ruledtabular}
\end{table}

\section{Numerical simulation details}
\label{sec:NumericalDetails}

The numerical calculations in Sec.~\ref{secSimulations} were carried out to integrate Eq.~\eqref{eq:lt-qfpe-d} with Eqs.~\eqref{eq:quantumLioiv}--\eqref{Xipq} in the quantum cases and Eq.~\eqref{heom_cl} with Eq.~\eqref{L_cl} in the classical cases, using a  fourth-order Runge--Kutta algorithm with the predictor--corrector method incorporated into the fourth-order Adams--Bashforth method and the fourth-order Adams--Moulton method.\cite{press1988numerical}  We set the time step $\delta t = 0.001$. Uniform meshes were employed to discretize the Wigner function with mesh sizes of $N_{q}= 64$ and $N_{p}=64$ in the $q$ and $p$ directions, respectively.

Other parameters for the isothermal and thermostatic processes are listed in Tables~\ref{table:SimIsothermal} and~\ref{table:SimThermostatic}, respectively.

\begin{table}[!t]
\caption{\label{table:SimIsothermal} Parameter values used for the simulations of the isothermal process. Here, $d x$ and $d p$ are the mesh sizes for position and momentum, respectively, in the Wigner space. The integers $N$ and $K$ are the cutoff numbers used in the LT-QFPE.}
\begin{ruledtabular}
\begin{tabular}{lcccdd}
& $ A $  & $ N $ & $ K $ &  \multicolumn{1}{c}{$d x$}  &  \multicolumn{1}{c}{$d p$}
\\
\hline
\multirow{3}{*}{Classical} & $0.5$ & $\cdots$ & $\cdots$ & 0.2 & 0.2
\\
& $1.0$ & $\cdots$ & $\cdots$ & 0.2 & 0.2
\\
& $1.5$ & $\cdots$ & $\cdots$ & 0.2 & 0.2
\\
\hline
\multirow{3}{*}{Quantum} & $0.5$ & $6$ & $3$ & 0.2 & 0.2
\\
& $1.0$ & $7$ & $3$ & 0.2 & 0.2 
\\
& $1.5$ & $8$ & $3$ & 0.25 & 0.35
\\
\hline
\end{tabular}
\end{ruledtabular}
\end{table}

\begin{table}[!t]
\caption{\label{table:SimThermostatic} Parameter values used for the simulations of the thermostatic process.  Here, $d x$ and $d p$ are the mesh sizes for position and momentum, respectively, in the Wigner space. The integers $N$ and $K$ are the cutoff numbers used in the LT-QFPE.}
\begin{ruledtabular}
\begin{tabular}{lcccdd}
& $ T_0 $ & $ N $ & $ K $ & \multicolumn{1}{c}{$d x$}  &  \multicolumn{1}{c}{$d p$} 
\\
\hline
\multirow{3}{*}{Classical} & $0.3$ & $\cdots$ & $\cdots$ & 0.2 & 0.2
\\
& $1.0$ & $\cdots$ & $\cdots$ & 0.25 & 0.25
\\
& $5.0$ & $\cdots$ & $\cdots$ & 0.3 & 0.4
\\
\hline
\multirow{3}{*}{Quantum} & $0.3$ & $7$ & $3$ & 0.25 & 0.3
\\
& $1.0$ & $7$ & $2$ & 0.3 & 0.45
\\
& ${5.0}$ & $7$ & $1$ & 0.3 & 0.6
\end{tabular}
\end{ruledtabular}\end{table}

\bibliography{references,tanimura_publist}

\end{document}